\documentclass{aa}  

\usepackage{graphicx}
\usepackage{txfonts}
\usepackage{hyperref}
\hypersetup{
    colorlinks=true,
    linkcolor=blue,
    filecolor=magenta,  
    citecolor=blue,
    urlcolor=cyan,
    pdfpagemode=FullScreen,
    }

\def\setsymbol#1#2{\expandafter\def\csname #1\endcsname{#2}}
\def\getsymbol#1{\csname #1\endcsname}

\newbox\tablebox    \newdimen\tablewidth
\def\leaderfil{\leaders\hbox to 5pt{\hss.\hss}\hfil}
\def\tablenote#1 #2\par{\begingroup \parindent=0.8em
    \abovedisplayshortskip=0pt\belowdisplayshortskip=0pt
    \noindent
    $$\hss\vbox{\hsize\tablewidth \hangindent=\parindent \hangafter=1 \noindent
    \hbox to \parindent{$^#1$\hss}\strut#2\strut\par}\hss$$
    \endgroup}

\def\L2{\ifmmode L_2\else $L_2$\fi}

\def\DeltaT{\ifmmode \Delta T\else $\Delta T$\fi}
\def\deltat{\ifmmode \Delta t\else $\Delta t$\fi}
\def\fknee{\ifmmode f_{\rm knee}\else $f_{\rm knee}$\fi}
\def\Fmax{\ifmmode F_{\rm max}\else $F_{\rm max}$\fi}
\def\solar{\ifmmode{\rm M}_{\mathord\odot}\else${\rm M}_{\mathord\odot}$\fi}
\def\Msolar{\ifmmode{\rm M}_{\mathord\odot}\else${\rm M}_{\mathord\odot}$\fi}
\def\Lsolar{\ifmmode{\rm L}_{\mathord\odot}\else${\rm L}_{\mathord\odot}$\fi}
\def\inv{\ifmmode^{-1}\else$^{-1}$\fi}
\def\mo{\ifmmode^{-1}\else$^{-1}$\fi}
\def\sup#1{\ifmmode ^{\rm #1}\else $^{\rm #1}$\fi}
\def\expo#1{\ifmmode \times 10^{#1}\else $\times 10^{#1}$\fi}
\def\,{\thinspace}
\def\lsim{\mathrel{\raise .4ex\hbox{\rlap{$<$}\lower 1.2ex\hbox{$\sim$}}}}
\def\gsim{\mathrel{\raise .4ex\hbox{\rlap{$>$}\lower 1.2ex\hbox{$\sim$}}}}

\def\simprop{\mathrel{\raise .4ex\hbox{\rlap{$\propto$}\lower 1.2ex\hbox{$\sim$}}}}
\def\deg{\ifmmode^\circ\else$^\circ$\fi}
\def\pdeg{\ifmmode $\setbox0=\hbox{$^{\circ}$}\rlap{\hskip.11\wd0 .}$^{\circ}
          \else \setbox0=\hbox{$^{\circ}$}\rlap{\hskip.11\wd0 .}$^{\circ}$\fi}
\def\arcs{\ifmmode {^{\scriptstyle\prime\prime}}
          \else $^{\scriptstyle\prime\prime}$\fi}
\def\arcm{\ifmmode {^{\scriptstyle\prime}}
          \else $^{\scriptstyle\prime}$\fi}
\newdimen\sa  \newdimen\sb
\def\parcs{\sa=.07em \sb=.03em
     \ifmmode \hbox{\rlap{.}}^{\scriptstyle\prime\kern -\sb\prime}\hbox{\kern -\sa}
     \else \rlap{.}$^{\scriptstyle\prime\kern -\sb\prime}$\kern -\sa\fi}
\def\parcm{\sa=.08em \sb=.03em
     \ifmmode \hbox{\rlap{.}\kern\sa}^{\scriptstyle\prime}\hbox{\kern-\sb}
     \else \rlap{.}\kern\sa$^{\scriptstyle\prime}$\kern-\sb\fi}
\def\ra[#1 #2 #3.#4]{#1\sup{h}#2\sup{m}#3\sup{s}\llap.#4}
\def\dec[#1 #2 #3.#4]{#1\deg#2\arcm#3\arcs\llap.#4}
\def\deco[#1 #2 #3]{#1\deg#2\arcm#3\arcs}
\def\rra[#1 #2]{#1\sup{h}#2\sup{m}}

\def\dots{\relax\ifmmode \ldots\else $\ldots$\fi}
\def\WHzsr{\ifmmode $W\,Hz\mo\,sr\mo$\else W\,Hz\mo\,sr\mo\fi}
\def\mHz{\ifmmode $\,mHz$\else \,mHz\fi}
\def\GHz{\ifmmode $\,GHz$\else \,GHz\fi}
\def\mKs{\ifmmode $\,mK\,s$^{1/2}\else \,mK\,s$^{1/2}$\fi}
\def\muKs{\ifmmode \,\mu$K\,s$^{1/2}\else \,$\mu$K\,s$^{1/2}$\fi}
\def\muKRJs{\ifmmode \,\mu$K$_{\rm RJ}$\,s$^{1/2}\else \,$\mu$K$_{\rm RJ}$\,s$^{1/2}$\fi}
\def\muKHz{\ifmmode \,\mu$K\,Hz$^{-1/2}\else \,$\mu$K\,Hz$^{-1/2}$\fi}
\def\MJysr{\ifmmode \,$MJy\,sr\mo$\else \,MJy\,sr\mo\fi}
\def\MJysrmK{\ifmmode \,$MJy\,sr\mo$\,mK$_{\rm CMB}\mo\else \,MJy\,sr\mo\,mK$_{\rm CMB}\mo$\fi}
\def\microns{\ifmmode \,\mu$m$\else \,$\mu$m\fi}

\def\muK{\ifmmode \,\mu$K$\else \,$\mu$\hbox{K}\fi}
\def\microK{\ifmmode \,\mu$K$\else \,$\mu$\hbox{K}\fi}
\def\muW{\ifmmode \,\mu$W$\else \,$\mu$\hbox{W}\fi}
\def\kms{\ifmmode $\,km\,s$^{-1}\else \,km\,s$^{-1}$\fi}
\def\kmsMpc{\ifmmode $\,\kms\,Mpc\mo$\else \,\kms\,Mpc\mo\fi}
\providecommand{\sorthelp}[1]{}

\usepackage{booktabs}
\newcommand{\bmat}[1]{\boldsymbol{ \mathsf {#1}}}
\newcommand{\bm}[1] {\boldsymbol{#1}}
\newcommand{\typeone}{\textsc{Type~1}} 
\newcommand{\typetwo}{\textsc{Type~2}} 
\newcommand{\typethree}{\textsc{Type~3}} 
\newcommand{\coj}{$J$\,=\,1\,$\to$\,0} 
\newcommand{\cojj}{$J$\,=\,2\,$\to$\,1} 
\newcommand{\cojjj}{$J$\,=\,3\,$\to$\,2} 
\newcommand{\kkms}{${\rm K_{RJ}\,km\,s^{-1}}$} 

\begin{document}

\title{Planck CO revisited: Improved CO line emission maps from Planck space mission observations}
\titlerunning{Planck CO maps revisited}

\author{Shamik Ghosh\thanks{\email{shamik@lbl.gov}} \inst{\ref{lbl}, \ref{cpb}}, Mathieu Remazeilles\thanks{\email{remazeilles@ifca.unican.es}} \inst{\ref{ifca}}, Jacques Delabrouille\thanks{\email{delabrouille@apc.in2p3.fr}} \inst{\ref{cpb},\ref{lbl}}}

\authorrunning{Ghosh, Remazeilles and Delabrouille}

\institute{Lawrence Berkeley National Laboratory, 1 Cyclotron Road, Berkeley, CA 94720, USA \label{lbl} \and Instituto de Fisica de Cantabria (CSIC-UC), Avda. los Castros s/n, 39005 Santander, Spain \label{ifca} \and CNRS-UCB International Research Laboratory, Centre Pierre Bin\'etruy, IRL 2007, CPB-IN2P3, Berkeley, CA 94720, USA \label{cpb}}

\date{Received xx,xx, 2023; accepted xx,xx, 2023}

\abstract{The Planck space mission has observed the first three rotational lines of emission of Galactic CO. Those maps, however, are either noisy, or contaminated by astrophysical emissions from different origin. We revisit those data products to deliver new full-sky CO maps with low astrophysical contamination and significantly enhanced noise properties. To that effect, a specific pipeline is designed to evaluate and postprocess the existing Planck Galactic CO maps. Specifically, we use an extension of the Generalized Needlet Internal Linear Combination method to extract multi-component astrophysical emissions from multi-frequency observations. Well characterized, clean CO full-sky maps at $10^\prime$ angular resolution are produced. These maps are made available to the scientific community and can be used to trace CO emission over the entire sky, and to generate sky simulations in preparation for future CMB observations.}

\maketitle

\section{Introduction}

The Galactic interstellar medium (ISM), which represents 10-15\% of the total mass of the Milky Way, is constituted of diffuse material in the Galactic plane and Galactic halo. Among its various components, cold molecular gas is concentrated in big molecular clouds. In such clouds, Carbon Monoxide (CO) is abundant, and  rotational lines of CO emission, at multiples of $\nu=115.27$~GHz, provide a convenient observable for tracing the distribution of cold molecular gas in the Milky Way. The lines of emission being located in frequency bands commonly used for Cosmic Microwave Background (CMB) observations, CO also is a foreground contaminant which must be cleaned-out for sensitive CMB observations. To that effect, reliable and high signal-to-noise ratio, full-sky tracers of CO emission would be useful for the preparation and analysis of upcoming sensitive CMB survey data over large fractions of high Galactic latitude sky.

Numerous ground-based observations of CO line emission have been performed in the past decades. Notably, a large  survey of the CO \coj\ line emission at low Galactic latitude was published by \citet{dame2001}, with a recent revision to extend the footprint of the survey \citep{2022ApJS..262....5D}. While many other observations exist on selected lines of sight or in limited regions of specific interest \citep{Hartmann1998, Ikeda1999, Fixsen1999, Magnani2000, Sawada2001, Mizuno2004, Wilson2005, Jackson2006, Bieging2010, Yoda2010, Bieging2011, Handa2012, Oka2012, Polychroni2012, Burton2013, Dempsey2013, Rigby2016, Colombo2019, Benedettini2020, Rennie2022, Dong2023, Park2023}, no spectroscopic survey of the full sky Galactic CO emission in various rotational lines exists as of now.

The Planck space mission, launched by ESA in 2009, was not designed to map CO emission. However, the exquisite sensitivity of the Planck detectors allowed for the release of full-sky observations of CO emission in the \coj, \cojj, and \cojjj\
rotational lines \citep{planck2013-p03a}. Two main data products have been released, dubbed `\typeone' and `\typetwo' CO maps. The \typeone\ maps make use of the different sensitivity to CO of detectors nominally in the same Planck frequency band. \typetwo\ maps assume a model of dust emission and fit, in each sky pixel, for a multi-component model of Galactic emission that includes CO lines, CMB and thermal dust. An additional data product, the `\typethree' CO map, assumes a constant line ratio between the various lines, and solves for a single CO map in a multi-component fit than maximizes the output CO map signal-to-noise ratio. CO products obtained with the Commander component separation method have also been obtained at HEALPix\footnote{\url{https://healpix.sourceforge.io/}} $N_{\rm side}=256$ for the three lines, and with $7.5^\prime$ angular resolution and $N_{\rm side}=2048$ for the \cojj\ line only, the latter assuming a fixed line ratio for the first three rotational lines  \citep{planck2014-a12}. 

Published Planck CO maps suffer from significant limitations. The \typeone\ maps, which use only single-frequency data to extract the CO, are in principle less prone to sky modeling errors, but the signal-to-noise ratio is low, in particular at high Galactic latitude where CO emission is faint. \typetwo\ maps, on the other hand, have reduced noise level, but are more prone to systematic errors due to variations of real sky emission scaling as compared to the model that is assumed, and are available only for the first two CO lines (\coj\ and \cojj). \typetwo\ maps are also delivered at lower angular resolution, $15^\prime$ for the two lines, instead of the native resolution of the Planck frequency channels, as is the case for \typeone\ maps. The \typethree\ map, which assumes fixed CO line ratio, cannot be trusted in regions of strong CO emission, where variations of the line ratio are clearly detected. Finally, the Commander maps are either at low HEALPix $N_{\rm side}$, or not available for all lines, and are also not obtained in a pipeline specifically optimized for mapping CO.

In this paper, we aim to extract improved maps of CO line emission for the three lines detected by Planck (\coj, \cojj\ and \cojjj), with reduced noise contamination, and taking into account the variation of the line ratio across the sky. To that effect, we post-process some of the published Planck CO map products, using a modified version of the Generalized Needlet Internal Linear Combination (GNILC) algorithm of \citet{Remazeilles:2011} in which we assume increasing priors as the signal-to-noise ratio decreases in the needlet domains being considered. This approach also produces maps at fixed high angular resolution, instead of the variable angular resolution that is obtained by the standard GNILC algorithm.\footnote{This work is based on observations obtained with Planck (\url{http://www.esa.int/Planck}), an ESA science mission with instruments and contributions directly funded by ESA Member States, NASA, and Canada.} 

The rest of this paper is organised as follows: In section \ref{sec:existing-maps}, we discuss and compare available Planck CO products and their limitations. Section~\ref{sec:reprocessing} presents the pipeline we use to produce reprocessed full-sky maps of the first three rotational lines of CO emission. Data products are presented in Section~\ref{sec:data-products} and discussed in Section \ref{sec:discussion}. We conclude in Section~\ref{sec:conclusion}.

\section{Existing Planck CO maps}
\label{sec:existing-maps}
In this work we only consider the Planck legacy data products for the CO line emissions. Here we discuss these legacy data products and their limitations.
\subsection{Single-frequency \typeone\ maps}

The Planck \typeone\ maps were generated for each CO line by forming a weighted linear combination of difference maps between the various pairs of bolometers associated with the single-frequency channel corresponding to the line's frequency, using the MILCA algorithm \citep{Hurier:2013,planck2013-p03a}. Taking into account that CMB and dust emissions do not vary significantly across bolometers within a specific frequency channel, whereas the CO line response depends much on the spectral bandpass of each bolometer, the MILCA weights were selected to deproject CMB and dust emissions (assuming both exhibit a uniform spectrum across the bolometers in thermodynamic units), while preserving the CO signal. 

The Planck \typeone\ CO maps consist of three full-sky maps, corresponding to the rotational transition lines \coj, \cojj\ and \cojjj. The angular resolution and HEALPix $N_{\rm side}$ of the maps are those of the original Planck frequency maps, per frequency channel. The single-channel methodology employed for \typeone\ CO maps effectively reduces the potential contamination from thermal dust, which could otherwise occur with multi-frequency combinations. However, it does come at the cost of a lower signal-to-noise ratio in contrast.

\begin{figure}
    \centering
    \includegraphics[width=0.47\textwidth]{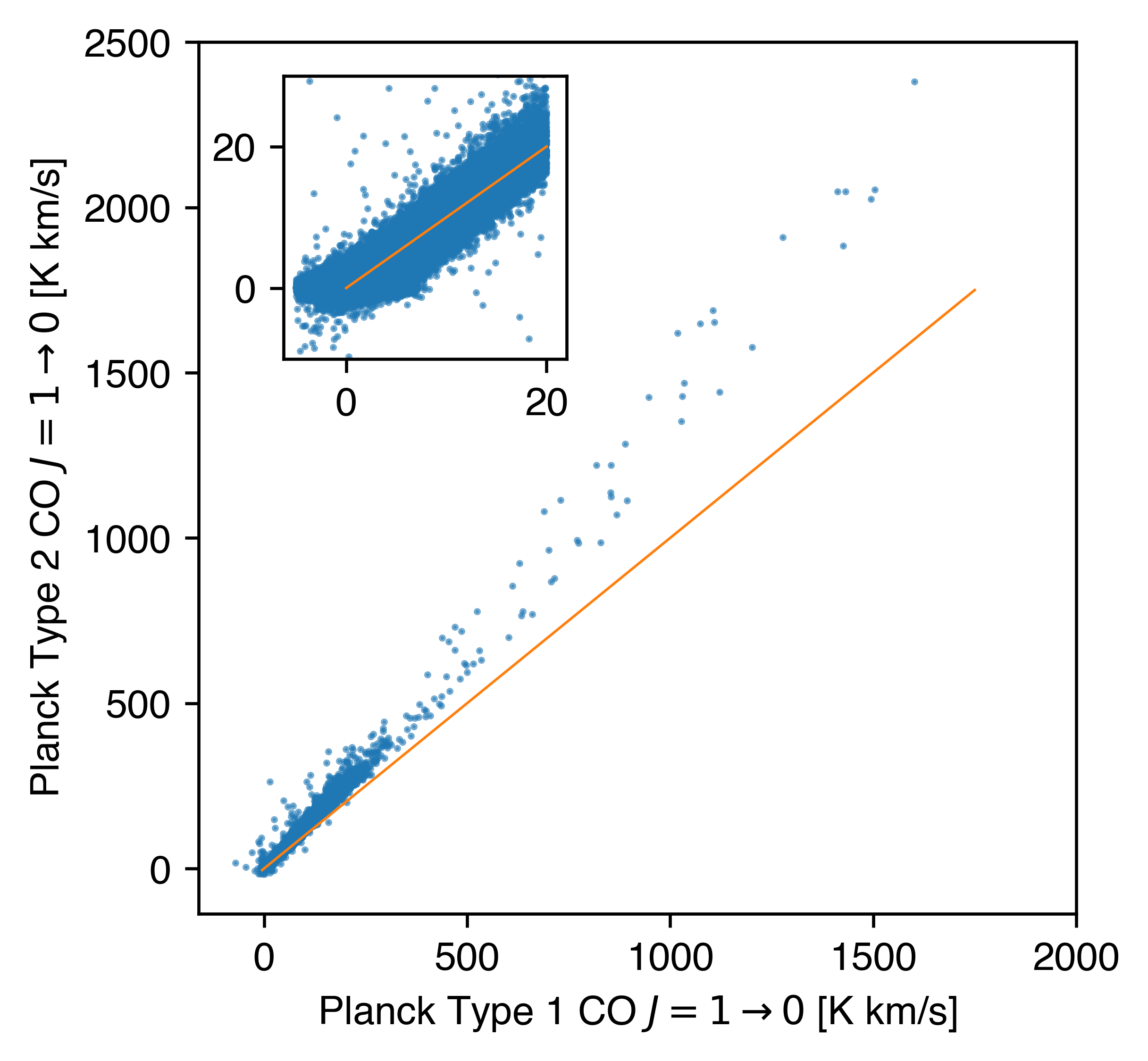}\\
    \includegraphics[width=0.47\textwidth]{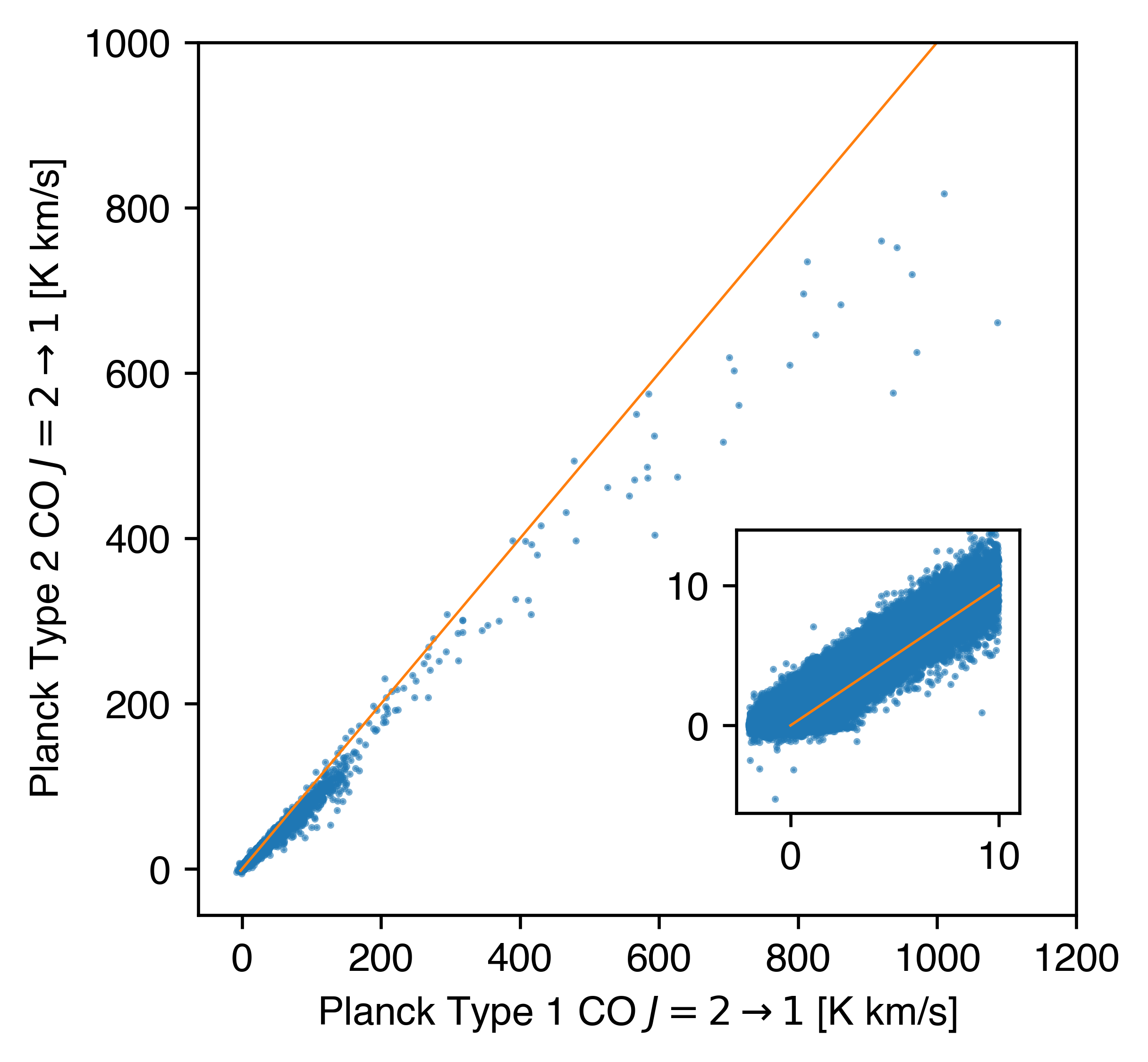}
    \caption{Scatter plot with pixel values of \typeone\ CO map as abscissa and pixel values of \typetwo\ CO maps as ordinate. We smooth all maps to 30 arcmin resolution and increase the HEALPix pixel size to $N_{\rm side}=256$ for this figure. The top figure shows the scatter for CO J=1$\rightarrow$0 Planck maps and the bottom figure for CO J=2$\rightarrow$1 maps. For large signals, the signal in \typetwo\ maps is higher than in \typeone\ maps for the J=1$\rightarrow$0 line, and lower for the J=2$\rightarrow$1 line. In the inset window we show that the low brightness pixels does not show obvious inconsistency. This is in agreement with the comparison made in \citet{planck2013-p03a}.}
    \label{fig:ty1-ty2-scatter}
\end{figure}

\subsection{Multi-frequency \typetwo\ maps}

The \typetwo\ maps have been obtained by applying a multi-component Generalized Least Square (GLS) filter \citep[Ruler algorithm;][]{planck2013-p06,planck2013-p03a} to multi-frequency Planck observations to disentangle CO emission from other astrophysical foregrounds. This requires the prior assumption of a parametric model for Galactic emission, used in the GLS mixing matrix. 

The Planck \typetwo\ CO maps consist of only two full-sky maps, corresponding to the lines \coj\ and \cojj. Both maps are produced at a common angular resolution of $15^\prime$, and HEALPix $N_{\rm side}$ of 2048. The multi-channel approach enhances the signal-to-noise ratio for \typetwo\ CO maps in comparison to \typeone\ CO maps. However, it is more prone to residual foreground contamination, stemming from thermal dust, free-free and radio source emissions that get partially projected into the final products by reason of potential modeling errors across the frequency channels.

Specifically, \typetwo\ CO maps are obtained by assuming negligible CO emission (as compared to dust) in the 353~GHz Planck channel, fitting a linear mixture of CO, dust, CMB and free-free emission in each pixel, with fixed emission laws for each of the extra components besides CO \citep[for details, see paragraph 2.2 of ][]{planck2013-p03a}. For free-free, the spectral index is assumed to be $\alpha_{\rm ff}=-2.15$ in brightness units, while the dust model assumes uniform dust temperature ($T_{\rm dust}=17$~K) and spectral index ($\beta_{\rm dust}=1.6$). Hence, for instance, the presence of a strong steep spectrum ($\alpha_{\rm source} < -2.15$) radio source would result in an over-estimate of the low frequency foreground at the line frequency, and thus an under-estimate of the CO emission. Similarly, dust emission with spectral parameters very different from what is assumed will result in positive or negative leakage of dust emission in the \typetwo\ maps. 

\subsection{Commander maps}
A \cojj\ map is obtained with Commander at resolution $7.5^\prime$ and $N_{\rm side}=2048$ \citep{planck2014-a12}. This map results from a global fit of all astrophysical components across Planck frequency channels. As such, 
it is more prone to modeling errors generating confusion between components than the \typeone\ products. It is challenging to make a component separation pipeline that is optimal for \emph{all} of the components present in the observations. The Commander method is quite flexible, but its implementation in \citet{planck2014-a12} was not specifically tuned for a reliable mapping of CO emission lines in the Planck data sets.

Maps obtained with Commander at $N_{\rm side}=256$, which do not fully exploit the angular resolution of the Planck observations, are not used in the reprocessing performed here.

\subsection{Limitations of existing maps}
\label{sec:limitations_of_existing_maps}

\begin{figure}
    \centering
    \includegraphics[width=0.45\textwidth]{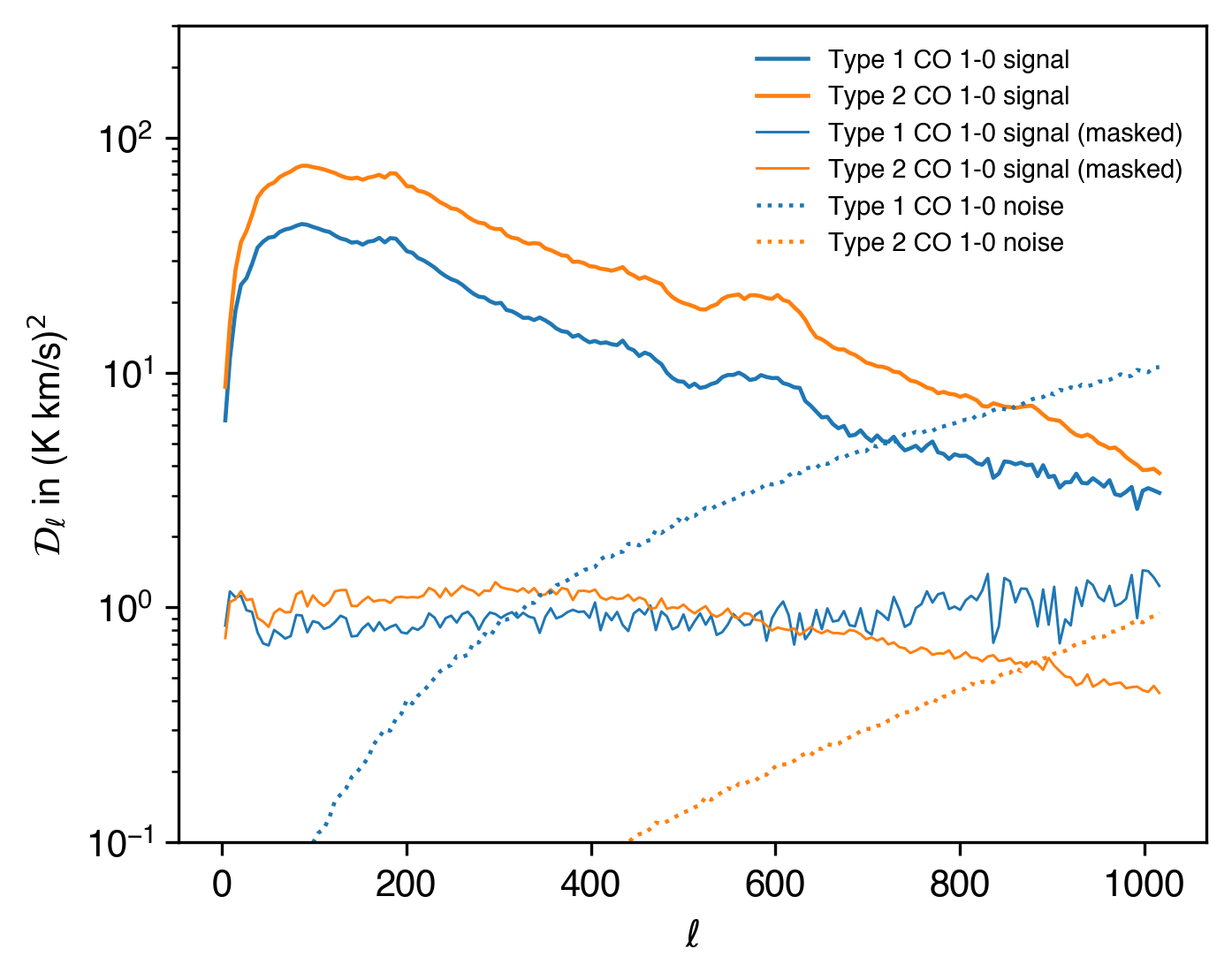}\\
    \includegraphics[width=0.45\textwidth]{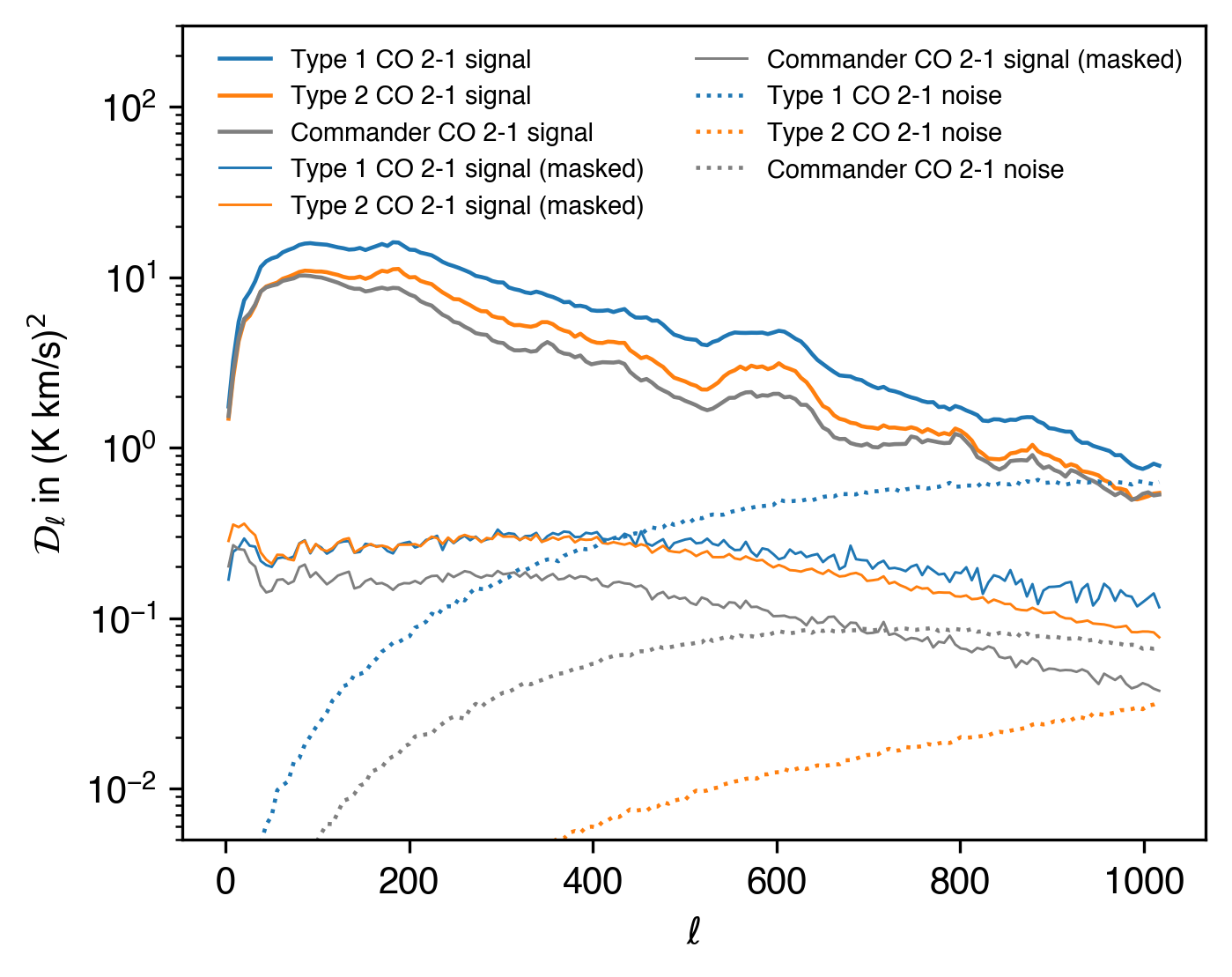}
    \caption{Power spectra comparison of \typeone\ and \typetwo\ CO \coj\ maps (top), and \cojj\ maps (bottom), with and without an attenuation mask to reduce the amplitude of emission in the brightest 2\% pixels (shown in the top panel of Figure \ref{fig:masks}). The \typeone\ \coj\ have lower power compared to the \typetwo\ map. For the CO \cojj\ line emission, the \typeone\ map has higher power compared to the \typetwo\ map. Both of them have higher power than the Commander CO \cojj\ map. The discrepancy between \typeone\ and \typetwo\ CO \cojj\ power spectra largely goes away with masking of the regions of strongest CO.}
    \label{fig:ty1-ty2-spectra}
\end{figure}

\begin{figure}[h]
    \centering
    \includegraphics[width=0.47\textwidth]{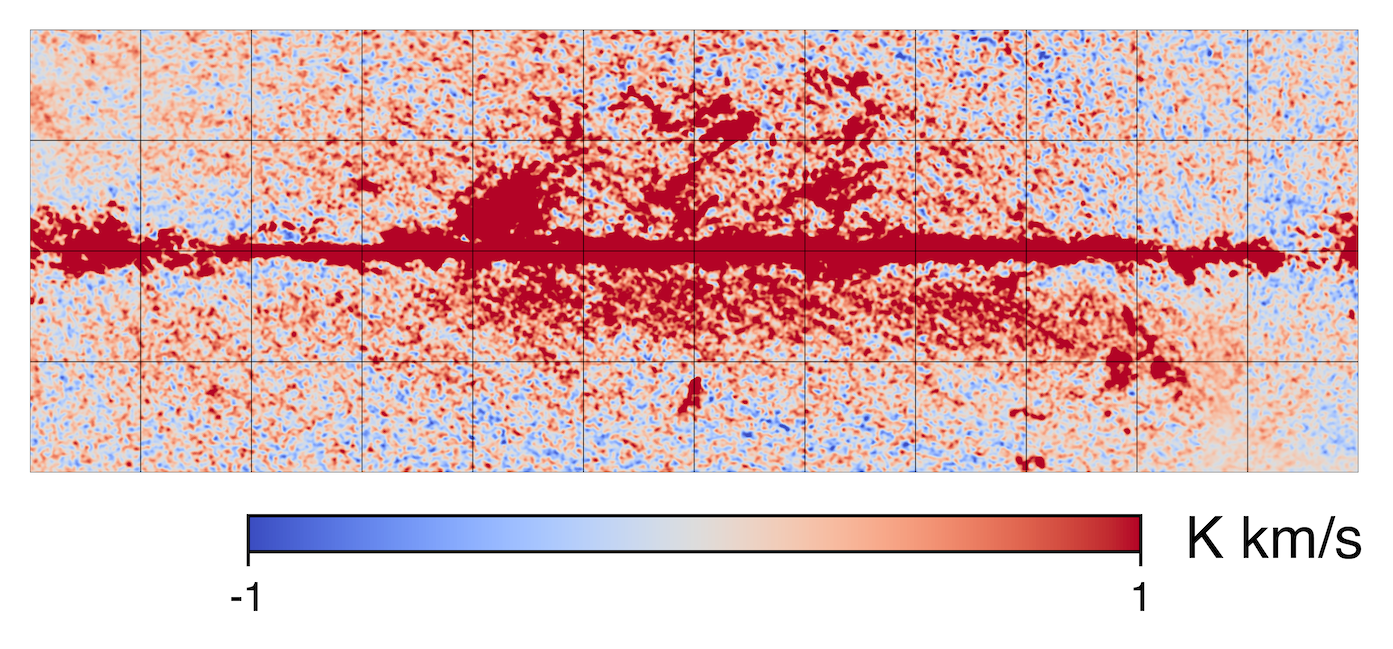}
    \caption{The \typeone\ CO \cojjj\ data product from Planck, shown here after smoothing to 30' resolution. We are showing a region $\pm 90^\circ$ in Galactic longitude and $\pm 30^\circ$ in Galactic latitude about the Galactic center. Ringing effects and systematic residuals are visible above and below the Galactic ridge.}
    \label{fig:type1-J3-2}
\end{figure}

Figure \ref{fig:ty1-ty2-scatter} plots \typetwo\ versus \typeone\ map values for all sky directions, showing a clear excess of \typetwo\ \coj\ and deficit of \typetwo\ \cojj\ emission as compared to \typeone\ products. This discrepancy is clearly visible for regions with large CO emission overall, but is not obviously seen for lower emission pixels. 

\begin{figure}[t]
    \centering
    \includegraphics[width=0.47\textwidth]{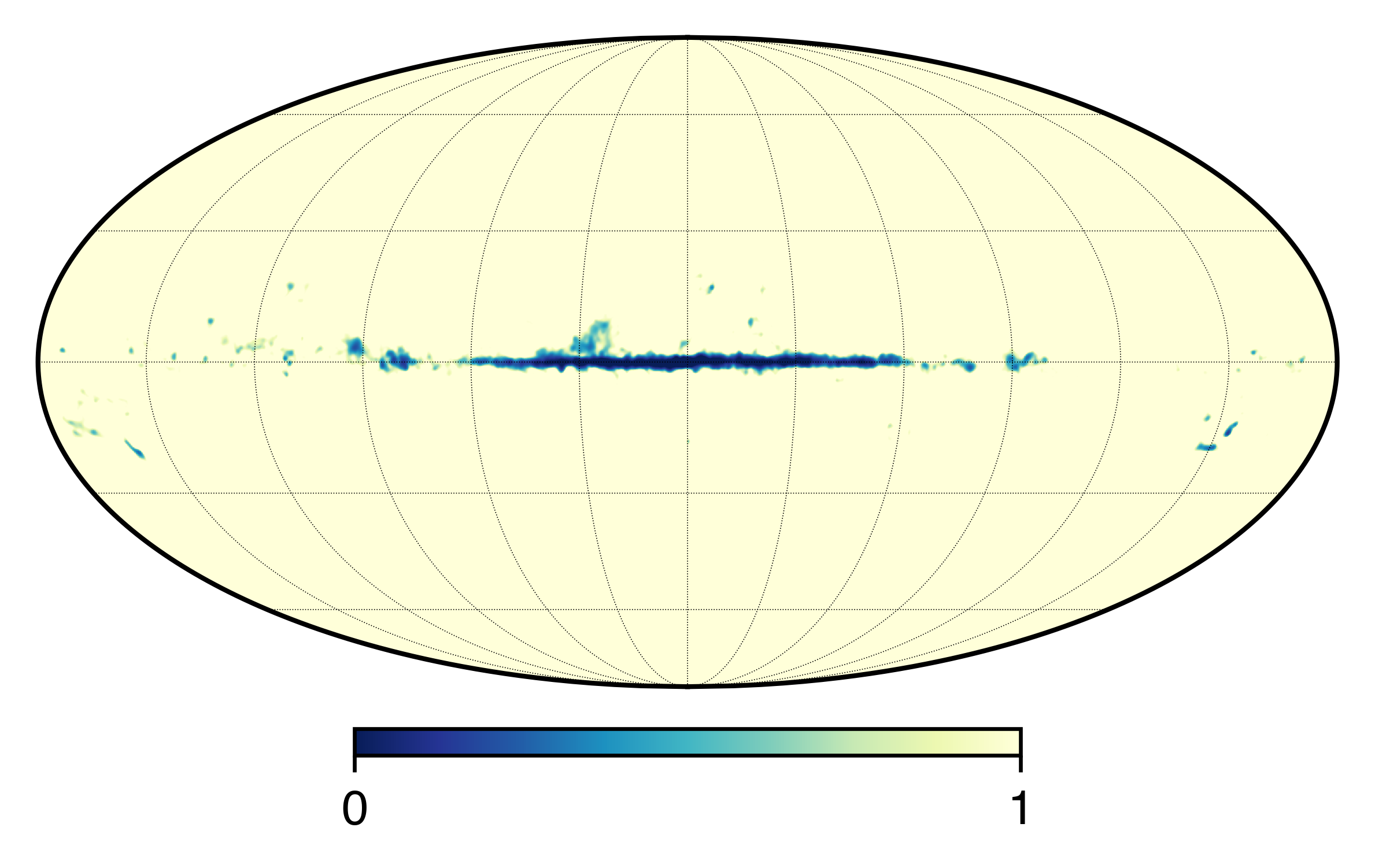}
    \includegraphics[width=0.48\textwidth]{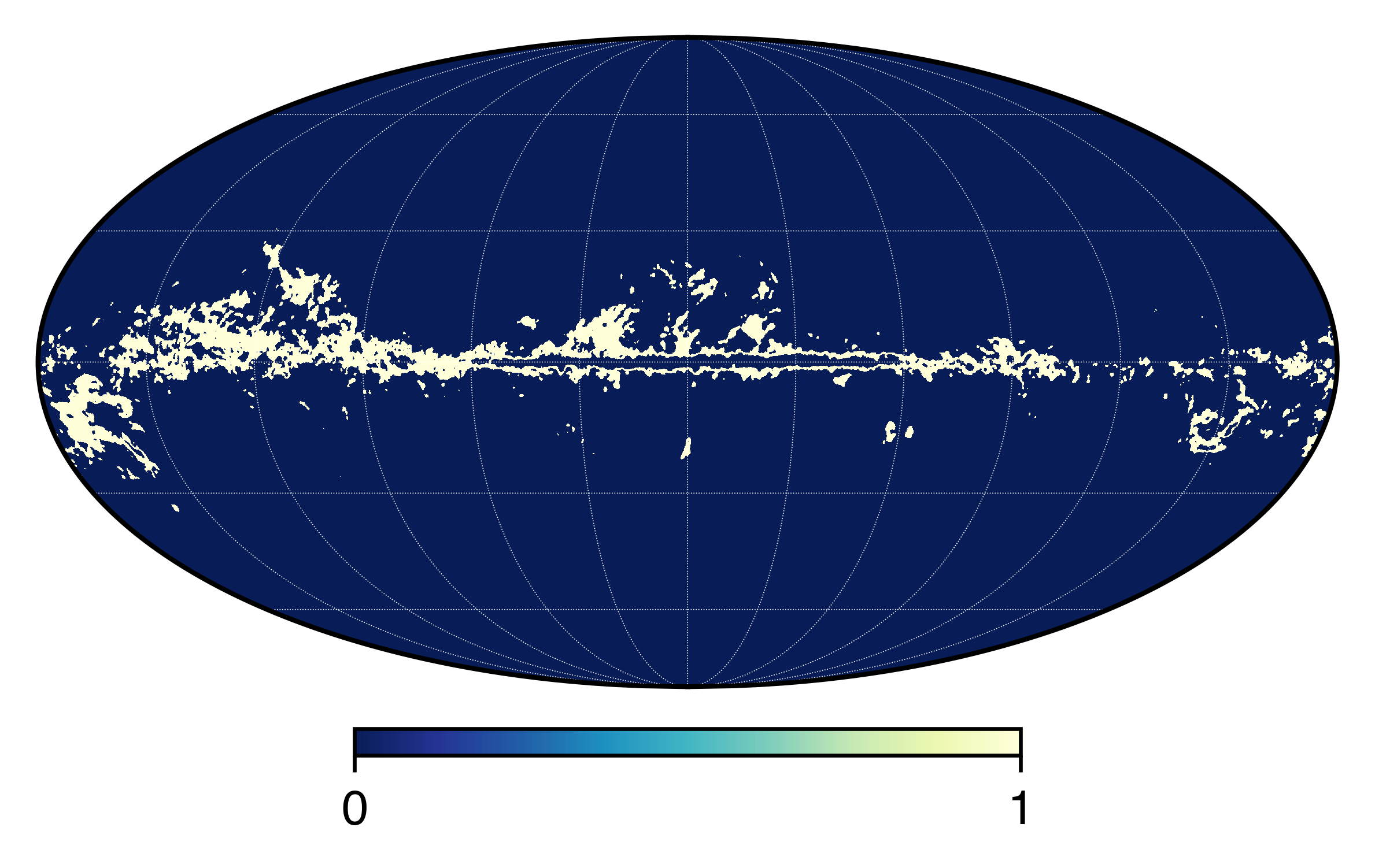}
    \caption{Top: Mask used to compute the power spectrum after excluding the brightest CO emission regions in the Galactic ridge. Bottom: Mask used for fitting the global CO mixing coefficients at low Galactic latitude the Galactic plane, obtained from the sky fraction observed in the \citet{dame2001} survey, excluding regions with CO emission larger than 3~\kkms. This results in the brightest 0.5\% of pixels (in the Galactic ridge) being masked.}
    \label{fig:masks}
\end{figure}

This calibration mismatch is also visible in a comparison of the power spectra of the \typeone\ and \typetwo\ maps (Figure \ref{fig:ty1-ty2-spectra}). 
The discrepancy between the \typeone\ and \typetwo\ maps is reduced when the power spectra are computed after masking the brightest sky regions using a small  mask of the Galactic ridge and of the brightest CO regions, displayed in the top panel of Figure \ref{fig:masks}. This is not the case, however, for the discrepancy between the Commander map and the other two, which is increased when the brightest regions are masked. 

As discussed in \citet{planck2013-p03a},  discrepancies between the \typeone\ and \typetwo\ maps likely originate from a mixture of contamination by $^{13}$CO and other diffuse Galactic emission residuals, and local inadequacy of the assumed emission spectral parameters for the generation of the \typetwo\ data products. Residual contamination from thermal Sunyaev-Zeldovich (SZ) emission is also evident in the \typetwo\ maps, with massive galaxy clusters such as Coma clearly visible in these maps, but not in \typeone\ maps. Since the \typeone\ data products are obtained from linear combinations of observations in the same frequency band, for which the contribution of all Galactic emissions other than CO lines are canceled, \typeone\ maps are more reliable tracers of the actual integrated CO emission (a linear combination of the main line and isotopologues), and less prone to contamination by other Galactic and extragalactic components.

The Commander map seems to lack power in low signal regions, and a visual inspection also shows significant contamination by SZ emission from galaxy clusters. As in addition the noise level is higher than that of the \typetwo\ map, the Commander map it is not used for the present work, as it brings little information that is not already present in \typeone\ and \typetwo\ maps.

Only the \typeone\ map is available for the \cojjj\ emission line. The central part of the map is displayed in Figure \ref{fig:type1-J3-2}. A visual inspection shows  large scale ringing around the Galactic plane, and residuals of scanning artefacts at higher Galactic latitude. Using comparisons with other maps, an attempt will be made at filtering some of these systematic effects in our improved Planck CO data products.

\section{Reprocessing pipeline}
\label{sec:reprocessing}


The different approaches that have been pursued to obtain \typeone, \typetwo, \typethree\ and Commander CO maps have advantages and drawbacks. As argued in the Planck CO paper \citep{planck2013-p03a}, \typeone\ maps are more reliable (in terms of separating CO from other emissions), but more noisy than the other products. This motivates an attempt at denoising those \typeone\ maps to improve their signal-to-noise ratio. To that effect, one can exploit the fact that the signal part in the \typeone\ CO maps for the various CO emission lines should be strongly correlated between maps, while the noise should be essentially independent. Similarly, the higher signal-to-noise ratio \typetwo\ maps can be used to help identify CO emission clouds locally in the \typeone\ maps. 

\subsection{Data model and methodology}
\label{sec:data_model}
For each of the \typeone\ and \typetwo\ maps (a total of five maps indexed by $\alpha=1,\dots,5$), we model the data $d_\alpha(p)$ as:
\begin{equation}
    d_\alpha(p) = s_\alpha(p) + n_\alpha(p),
\end{equation}
where $p$ denotes sky pixel, $s_\alpha(p)$ is the total sky signal in map $\alpha$ (mostly CO line emission, but possibly also some residuals of continuum Galactic emission, in particular for the \typetwo\ maps) and $n_\alpha(p)$ denotes the noise contamination. In vector format, this can be recast, for each pixel, as
\begin{equation}
    \bm{d}(p) = \bm{s}(p) + \bm{n}(p),
\end{equation}
where all vectors are 5-dimensional.
In the discussion that follows we will not explicitly write the pixel dependence, and unless explicitly stated all data vectors and matrices have a pixel dependence that is hidden.

The signal, $\bm{s}$, which is mostly due to emission from the same CO molecular clouds, is strongly correlated between these maps. The key idea for denoising is that we can write the signal in each map as a linear combination of $N_s$ 
templates (which may or may not be due solely to CO emission), $\bm{t}$, that trace the sky signal:
\begin{equation}
    s_\alpha = \sum_{i=1}^{N_s} A_{\alpha i} t_i, 
    \label{eq:signal_template}
\end{equation}
or 
\begin{equation}
    \bm{s} = \bmat A \bm{t}.
\end{equation}
In general, for $N_{ch}$ input maps and $N_s$ signal templates, $\bmat A$ denotes a $N_{ch} \times N_s$ mixing matrix. If $N_s$ is strictly lower than $N_{ch}$ (which is expected to be the case, considering that the signal is highly correlated among the various CO input maps), we can project the data on the subspace spanned by the $N_s$ templates that are required to model the signal, and hence get rid of part of the noise contamination. Doing this with simultaneous localization in harmonic space and pixel domain is the main idea behind GNILC \citep{Remazeilles:2011}, which we use for the present work, but with a modification that we describe next.\footnote{This way of modeling `multidimensional' foreground components is also used in recent implementations of the SMICA component separation method \citep{cardoso2008}.}

For some applications indeed, one of the drawback of the standard GNILC procedure is that when the dimension $N_s$ of the foreground subspace is zero in some needlet domain (i.e. in some needlet band, in some region of sky), the data for that needlet domain is completely discarded (projection onto a zero-dimensional subspace). This results in varying angular resolution over the sky, which complicates the map characterization and its scientific exploitation. This can be alleviated by imposing that the dimension of the foreground subspace be always at least equal to one, and more if the data requires so. If, however, one keeps the dimension corresponding to the largest eigenvalue of the covariance matrix, in regions of low signal one would keep the dimension with the highest noise excursion. 
Hence, in practice, when the GNILC procedure would suggest discarding the observations as too noisy (i.e. projecting on a zero-dimensional subspace), we use instead a one-dimensional prior for the signal subspace, and combine all observations using a least-square estimate of the signal assuming the corresponding signal color. This amounts to co-adding all observations with weights that maximize the signal-to-noise ratio for a pre-determined signal color, and then rescaling the resulting single CO map to inject its contribution in each individual line map. 


\begin{figure*}
    \centering
    \includegraphics[width=0.32\textwidth]{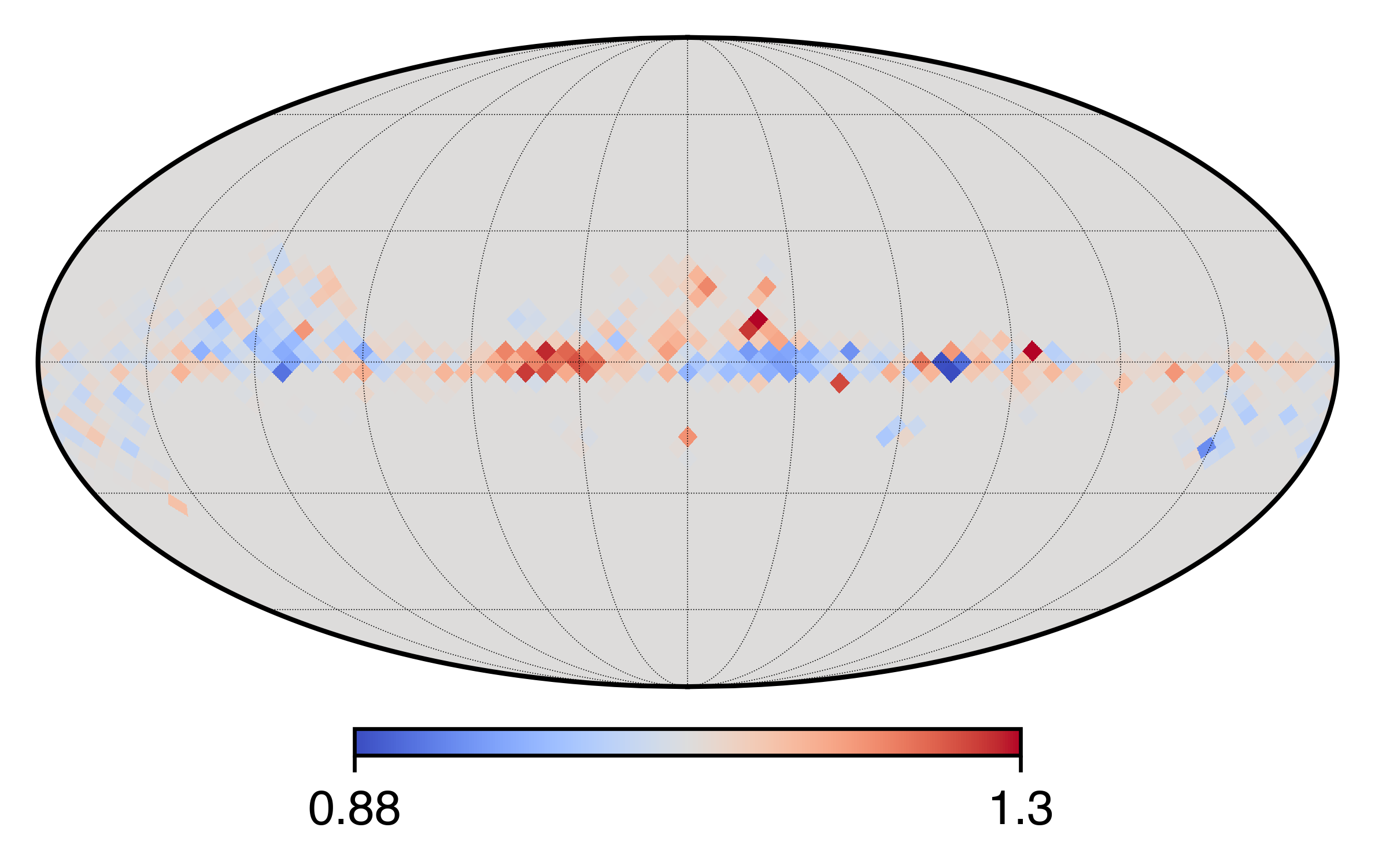}
    \includegraphics[width=0.32\textwidth]{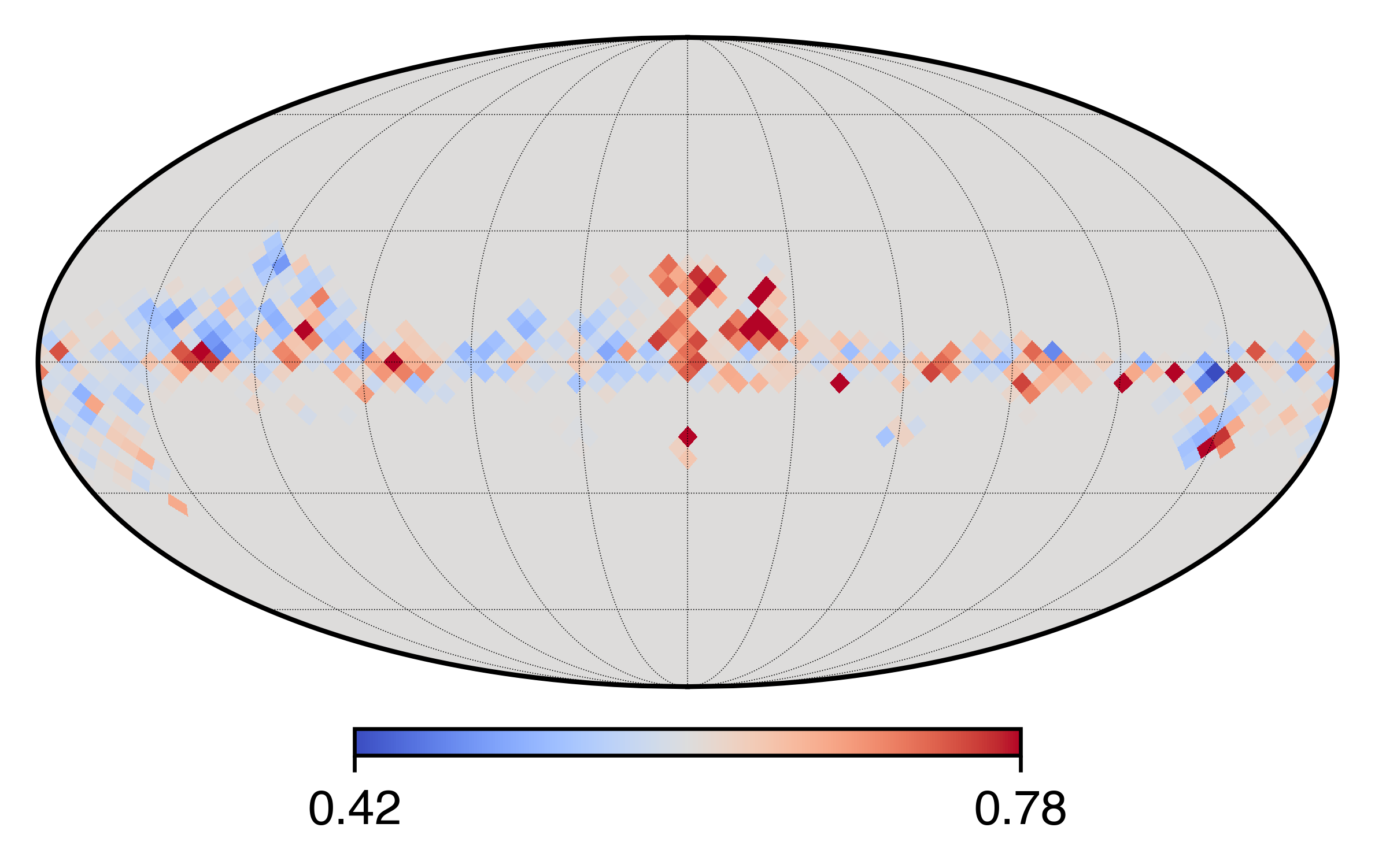} 
    \includegraphics[width=0.32\textwidth]{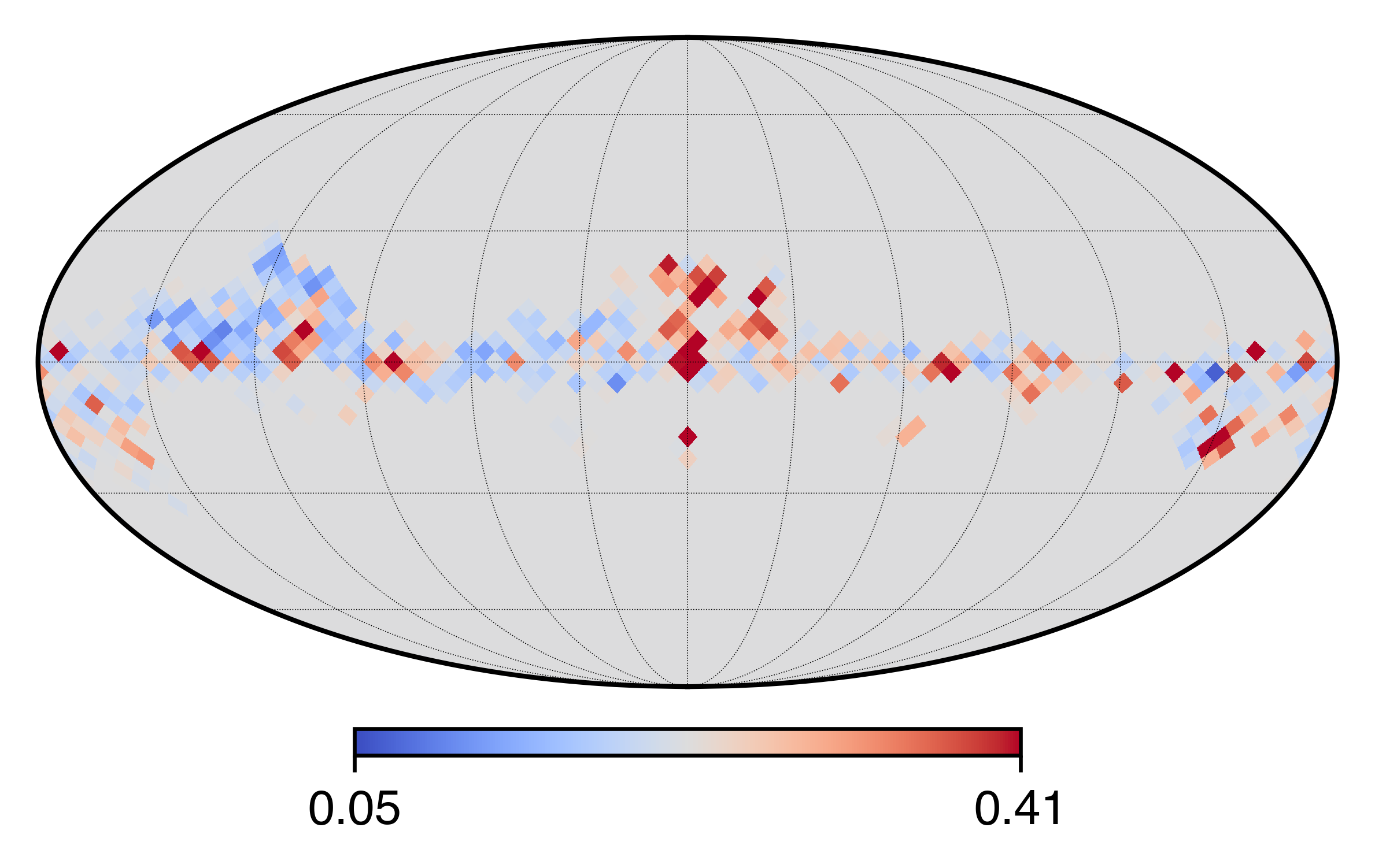}\\
    \includegraphics[width=0.32\textwidth]{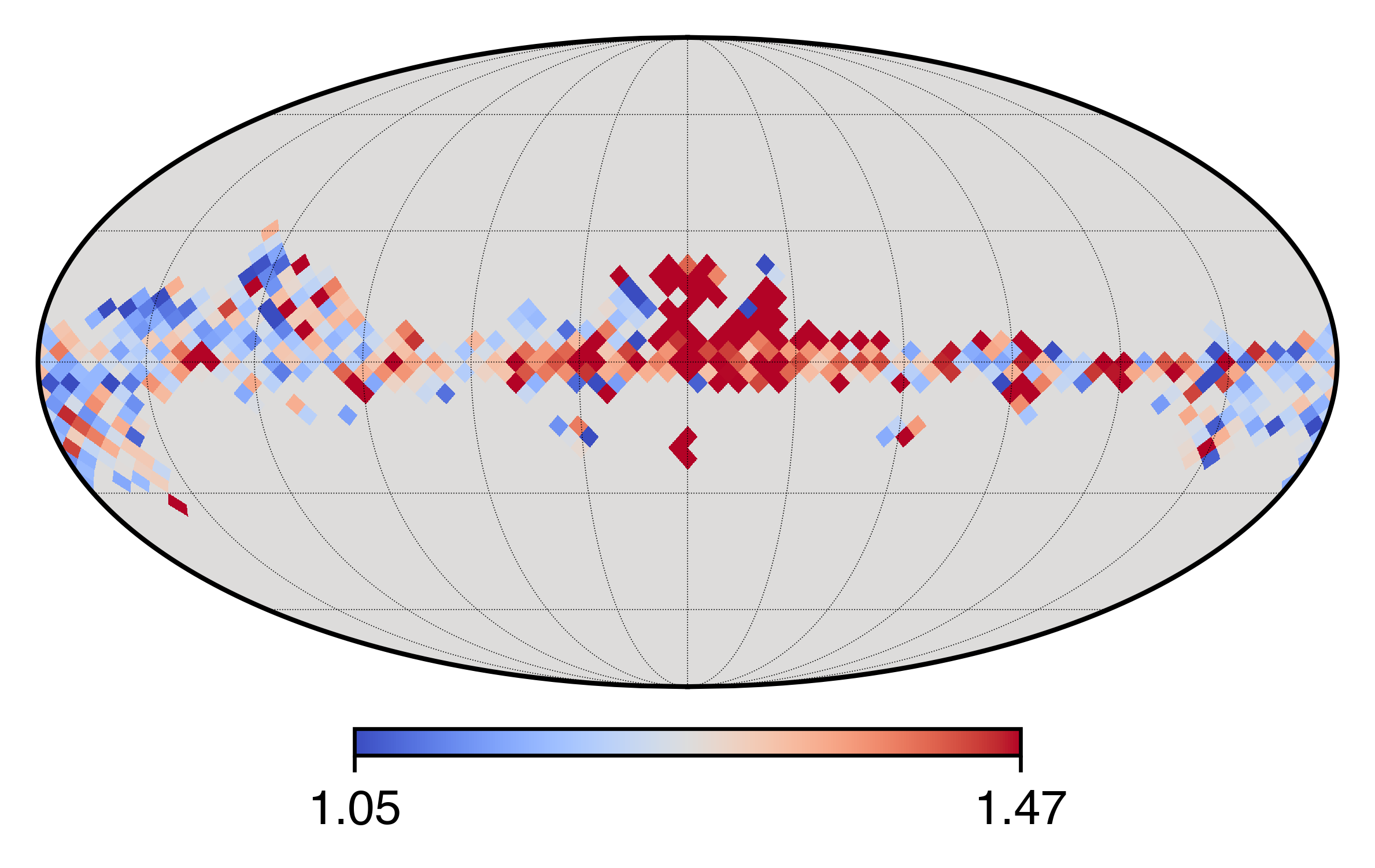}
    \includegraphics[width=0.32\textwidth]{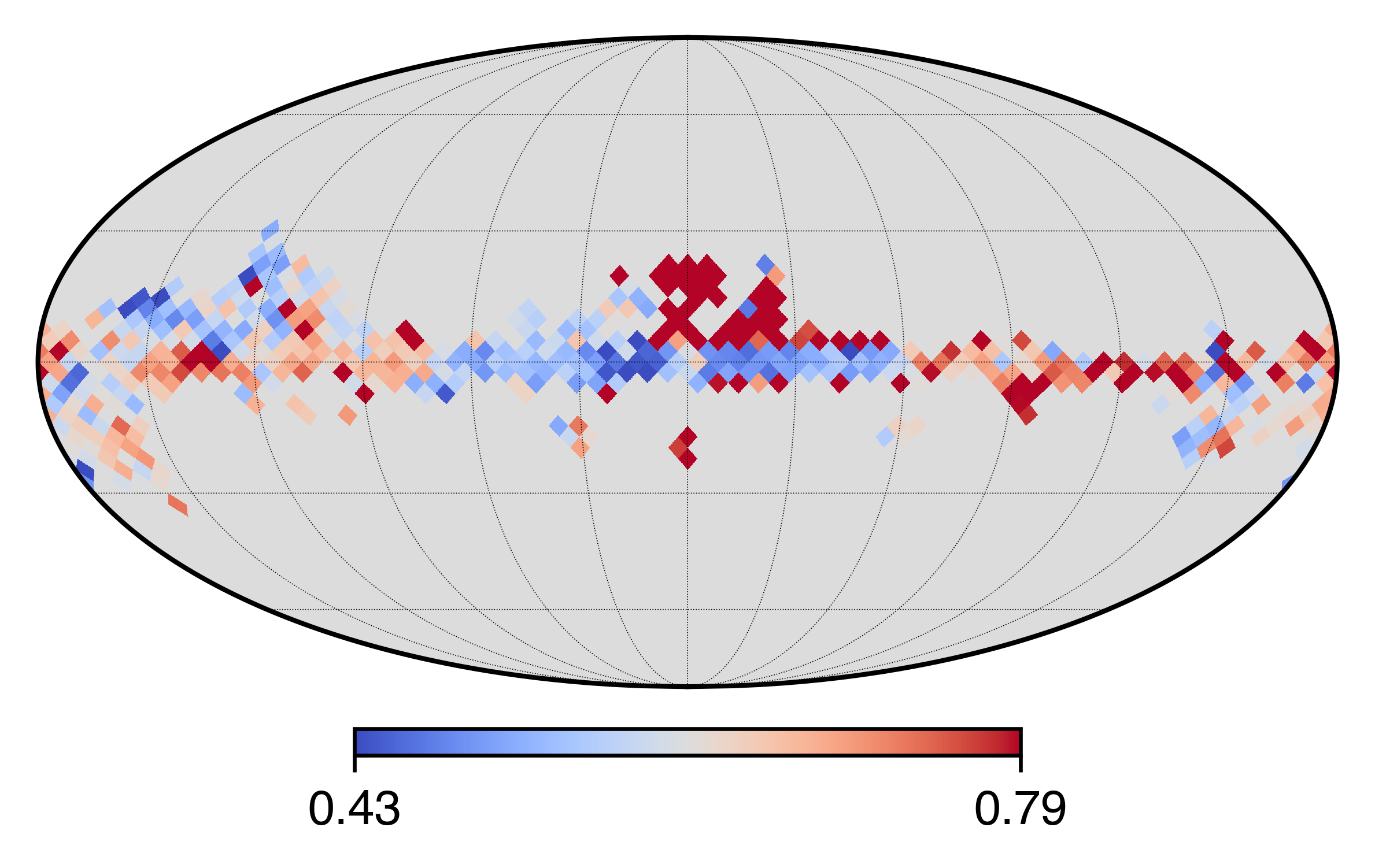}
    \caption{Maps of calibration obtained by fitting Planck \typeone\ (top row, from \coj\ to \cojjj, left to right) and \typetwo\ (bottom row, from \coj\ to \cojj, left to right) maps against Dame \coj\ map in $N_{\rm side}=16$ super pixels. We perform a fit in any super pixel if there are 50 or more original pixels at $N_{\rm side}=512$ with high signal to noise ratio in the Dame map. For all super pixels where this condition is not met, we assume the mean calibration given in Table \ref{tab:calibration}. These maps illustrate the variation over the sky of the ratio between the CO detected in Planck \typeone\ and \typetwo\ maps, and the CO detected in the Dame \coj\ map.}
    \label{fig:calibration-maps}
\end{figure*}

\subsection{Overview of the pipeline}

The main steps of the processing pipeline can be summarized as follows:
\begin{enumerate}
\item \emph{Preprocessing}: The input data to the pipeline are the set of 5 published Planck \typeone\ and \typetwo\ CO maps, smoothed to a common $10^\prime$ angular resolution. We mask or saturate some of the brightest pixels to avoid ringing effects polluting the neighborhood when we perform harmonic transforms for the needlet decomposition; We also mask in \typetwo\ maps 186 bright galaxy clusters, the thermal SZ emission of which leaked in the Planck data products (this leakage is clearly visible for the Coma cluster, for instance). We inject noise from null maps (i.e. half-difference of half-ring Planck CO maps) in the masked regions, to preserve the local statistics of the Planck maps noise.
\item \emph{Calibration}: We compute a best-estimate `CO mixing vector' for the set of five maps, by calibrating each map against the \coj\ map of \citet{dame2001}, reprojected into a Healpix map and available on the LAMBDA website;
This provides us with a prior for the signal subspace (i.e. the five coefficients $a_{\alpha} = A_{\alpha 1}$ for one single template in Equation \ref{eq:signal_template}), in needlet domains where no signal is detected above the noise in the auto- and cross-covariance of the input maps. 
\item \emph{Extended GNILC pipeline (named "xGNILC")}: We compute needlet coefficients for all five input CO maps, and implement a modified GNILC algorithm, in which a Generalised Least Square (GLS) solution assuming the CO scalings from the previous step is used in needlet domains where the Akaike Information Criterion (AIC) suggests a zero-dimensional GNILC signal subspace. We obtain output maps for each of the \typeone\ and \typetwo\ line maps; Only outputs corresponding to the \typeone\ inputs are kept in our final set of CO products. For all lines, the signal that has been subtracted from the input Planck CO maps in the masking or saturation process in step 1 is re-injected in the output maps.
\item \emph{Postprocessing}: We postprocess the output xGNILC \cojjj\ map to subtract large-scale systematics originating from the input Planck \typeone\ \cojjj\ product. 
\end{enumerate}
Details about each of these steps are provided in the next subsections.

\subsection{Preprocessing}
\label{sec:preproc}

Although in principle the \cojj\ and \cojjj\ maps allow for an angular resolution of the order of $5^\prime$, the signal-to-noise ratio becomes very low on the smallest scales. The output data products will be maps at $N_{\rm side}$ of 1024, with $\ell_{\rm max}$ of 2048. We choose a common resolution of $10^\prime$, but we `apodize' the Gaussian beam to smoothly go to zero by $\ell_{\rm max}$ of 2048, as shown in Fig. \ref{fig:needlets}. In the text that follows, we denote this `apodized' beam simply as $10^\prime$ beam. Hence, as a first step, we smooth all maps to a common $10^\prime$ angular resolution (only little information can be extracted from the observations on smaller scales in practice---and mostly in Galactic regions where other data from spectroscopic observations of the CO lines exist).

Next, we note that the Planck CO maps are very bright towards the Galactic center and towards a few compact regions of emission, mostly near the Galactic plane. These small regions of bright emission tend to produce ringing features when maps are decomposed into needlets. To avoid this unwanted feature we identify these regions, mask and fill them. The exact process goes as follows:
\begin{enumerate}
    \item Identify region of very bright CO emission and determine a brightness threshold in ${\rm K_{RJ}\, km \, s}^{-1}$ that is required to create a clipping mask for the region.
    \item Mask and set the masked pixels to the threshold value. Perform a median filter on this map, 
    with the median computed over superpixels at $N_{\rm side}=256$, and smooth the median-filtered map with a $15^\prime$ beam.
    \item Use a $1^\circ$ apodized mask to mask the original map and fill the gaps with the smoothed median-filtered map obtained above.
\end{enumerate}

We save the difference between the maps obtained in this way and the initial maps, for re-injection of these strong signals in the final data products. 

The Planck \typetwo\ maps are contaminated by thermal SZ (tSZ) emission. To produce a mask of tSZ emission we use a simulation of known tSZ sources from a recent version (v2.3.2) of the Planck Sky Model \citep{delabrouille2012}. We produce a mask where we set the $6.5\times 10^{-3}$ \% brightest pixels of the simulated tSZ map to zero. We are careful not to mask anything with strong CO signal, and to that effect exclude from the masked pixels any pixel in the Galactic plane where either the Dame CO map has a value greater than 0.5 ${\rm K_{RJ}\, km \, s}^{-1}$, or where in the faint clouds away from the Galactic plane the Planck \typetwo\ CO \cojj\ map is larger than 0.25 ${\rm K_{RJ}\, km \, s}^{-1}$. This excludes any masking that may be too close to any region containing strong CO signal.\footnote{The thresholds, while arguably a bit \emph{ad hoc}, are selected so that no strong CO signal is masked to cut-out a nearby faint tSZ cluster.} We additionally mask circular regions with radii $0.35^\circ$ and $0.55^\circ$ around the center of the Virgo and Coma clusters, respectively. This mask is finally apodized with a $0.6^\circ$ cosine taper. We mask and fill the Planck \typetwo\ CO maps with the corresponding Planck \typetwo\ null maps using the apodized tSZ emission mask.

\subsection{Calibration}
\label{sec:calibration}

Calibration against the \citet{dame2001} map provides us with estimates of the ratio of CO emission in each Planck CO map to the CO emission in the Dame map. To perform this calibration, we first smooth all maps to a common $30^\prime$ angular resolution, and downgrade them to HEALPix $N_{\rm side}=512$. We also generate, for all Planck products, corresponding statistical uncertainty maps downgraded to $N_{\rm side}=512$. We then consider two options for performing a global calibration.

\begin{enumerate}
    \item In option 1, we select a sky region in the Dame map using the mask displayed in the bottom panel of Figure \ref{fig:masks}. For each Planck CO map, we then compute a global best fit for a multiplicative factor $\bar{a}_{\alpha}$ by sampling a Gaussian likelihood with a uniform prior of $0 \le {a}_{\alpha} \le2.5$. For this fit, we assume a model in which ${\rm CO}_{{\textsc{Type X}}}^J = {a_{\alpha}} \times {\rm CO}_{\textsc{Dame}}^{J=1\to 0}$, where $X \in \{ 1,2 \}$ and $J$ indexes the CO line.    
    \item For option 2, we repeat the same procedure, but using  a fitting model with an offset, ${\rm CO}_{{\textsc{Type X}}}^J = {a_{\alpha}} \times {\rm CO}_{\textsc{Dame}}^{J=1\to 0} + K_{{\textsc{Type X}}}^J$. We assume for the offset a flat prior,  
    $K_{{\textsc{Type X}}}^J \in [-10,10]$~\kkms.
\end{enumerate}
We also investigate how the calibration varies across the sky. To that effect, we repeat the calibration after dividing the calibration region in sub-regions of equal sky area, obtained by sorting first the pixels used in the calibration in three equal-area Galactic latitude regions, and then each latitude region in 10 equal-area longitude regions (for a total of 30 sub-regions). We perform the calibration independently in each of those regions. We then compute the mean, the median, and the standard deviation of the 30 calibration coefficients obtained in this way.

All calibration results are reported in Table \ref{tab:calibration}, where they are also compared to expected values derived from the calibrations performed in the original Planck publication. We find reasonable consistency between the various results, and choose the coefficients obtained as the mean of 30 calibrations in different regions as our fiducial $\bar{a}_{\alpha}$ (from the fit with an offset). We use those values for the GLS solution in needlet domains of faint CO emission in the extended GNILC pipeline described later.

\begin{table*}
\caption{Best-fit estimates for the global scaling parameter $a_{\alpha}$, for various calibration methods. Statistical errors for the global fits are not reported, as they are very small (of the order of $10^{-4}$ to $10^{-3}$), and not representative of actual uncertainties given the variation of the calibration coefficients over the sky. The two Planck values reported for \typeone\ \coj\ are from a model with $^{13}{\rm CO}/^{12}{\rm CO}\simeq 0.2$ and a $0.53 \times {^{13}{\rm CO}}$ contamination estimate in the \typeone\ \coj\ map, and from a fit performed by the Planck team over selected sky regions.}
\label{tab:calibration}      
\centering                                      
\begin{tabular}{|c|ccc|ccc|c|}         
\toprule                       
Planck map & Global fit & Mean of 30 & Median of 30 & Global fit & Mean of 30 &  Median of 30 & Planck \\   
(type and line) & (with offset) & (with offset) & (with offset) & (no offset) & (no offset) & (no offset) & (estimate) \\
\midrule
\typeone\ \coj   &  $1.12$  &  $1.09 \pm 0.07$  & $1.08$  &  $1.21$  & $1.19 \pm 0.09$ & $1.20$  & 1.11 / 1.16 \\ 
\typeone\ \cojj  &  $0.61$  &  $0.60 \pm 0.06$  & $0.60$  &  $0.60$  & $0.59 \pm 0.06$ & $0.58$  & \\
\typeone\ \cojjj &  $0.21$  &  $0.23 \pm 0.06$  & $0.22$  &  $0.19$  & $0.20 \pm 0.04$ & $0.19$  & \\
\typetwo\ \coj   &  $1.26$  &  $1.26 \pm 0.07$  & $1.25$  &  $1.24$  & $1.24 \pm 0.08$ & $1.23$  & 1.2\\
\typetwo\ \cojj  &  $0.61$  &  $0.61 \pm 0.06$  & $0.62$  &  $0.67$  & $0.67 \pm 0.09$ & $0.66$  & \\
\bottomrule                                           
\end{tabular}
\end{table*}

As a final investigation of the calibration of Planck CO products relative to the \coj\ Dame map, we repeat the calibration locally over the entire sky, in local superpixels corresponding to HEALPix $N_{\rm side}=16$. For these local fits, performed for all sky superpixels where Dame data is present, we assume for ${a_{\alpha}}$ a Gaussian prior with mean and standard deviation values from the 30-regions analysis. Resulting calibration maps are displayed in Figure \ref{fig:calibration-maps}.  
 

\subsection{An extended GNILC pipeline}
The extended GNILC pipeline (dubbed xGNILC in the following) used in the present work has four main parts performed consecutively: 1 - the needlet decomposition; 2 - the determination of the CO subspace (in each sky pixel, for each needlet scale);  3 - the projection of the input data into final needlet coefficient maps using either a multidimensional ILC or a GLS solution (the latter where the dimension of the CO subspace is estimated to be null, i.e. $N_s = 0$, in step 2); and 4 - the recombination of the needlet maps into five processed maps. We describe each of these steps in more detail below.

\begin{figure}
    \centering
    \includegraphics[width=0.48\textwidth]{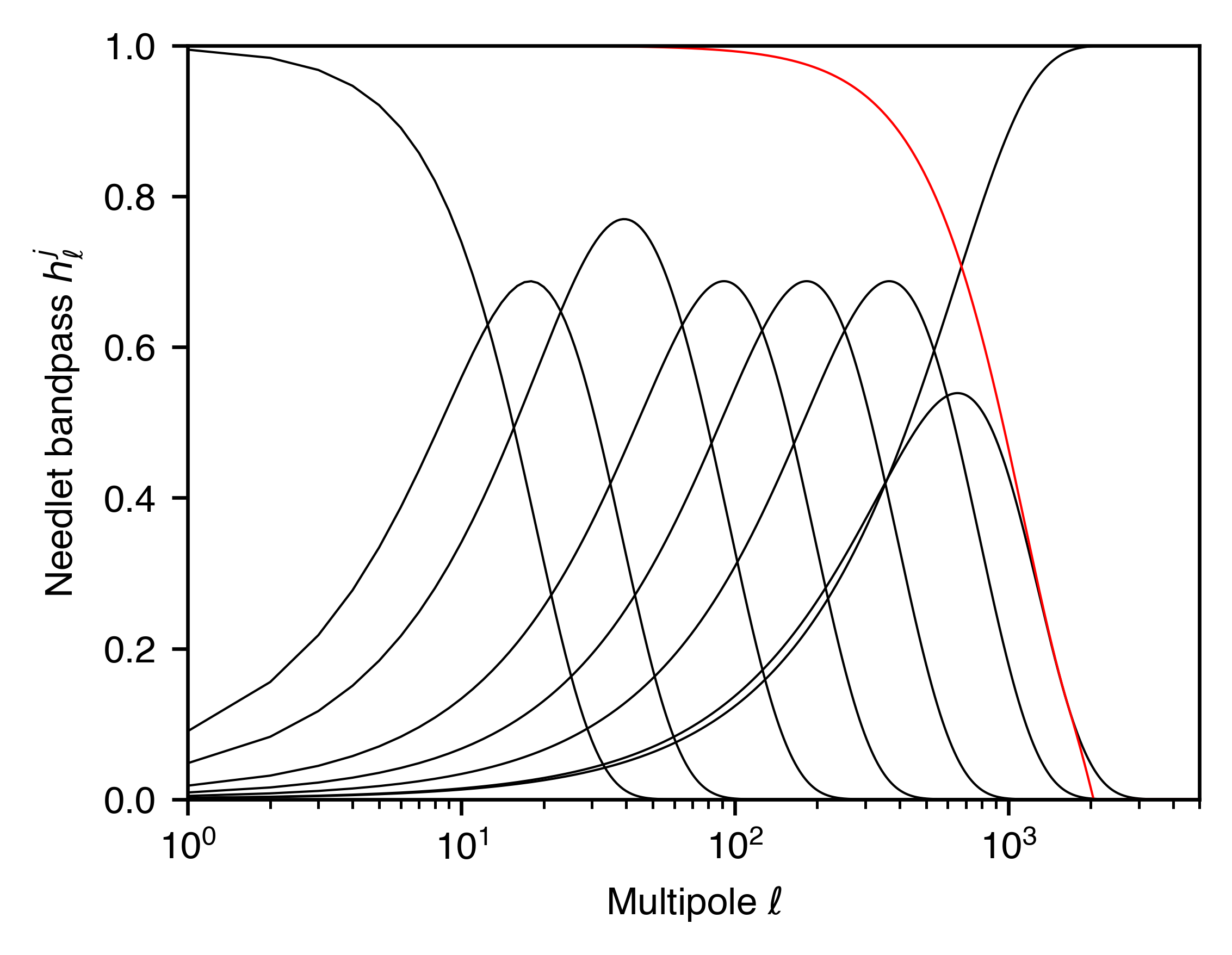}
    \caption{A plot of the needlet bandpass windows, $h_\ell^j$, showing the selection weights in harmonic space (plotted in black). The `apodized' beam transfer function of the output map resolution of $10^\prime$ is also shown in red. The end of the beam transfer function is modified to smoothly go to zero at $\ell_{\rm max}=2048$.}
    \label{fig:needlets}
\end{figure}
\subsubsection{Needlet decomposition}
The so-called `Needlets' are a kind of spherical wavelets, which allow us to compute statistics and perform map combinations locally in harmonic and in pixel spaces simultaneously \citep{Narcowich2006}. The originally introduced needlets have  spectral band shapes specifically tailored for good localization. In practice, applications for CMB data analysis \citep[e.g.][]{Pietrobon2008, Fay2008, delabrouille2009, Geller2008, Scodeller2011} have used various types of needlet bandpass windows, adapted to the angular resolution of the observations and the variability of the signals of interest and of the noise across the sky and in harmonic domain. For this work we use needlets which are defined in harmonic space with 8 Gaussian shaped bandpasses $h_\ell^j$, such that $\sum_j (h_\ell^j)^2 = 1$. These bandpasses are written as:
\begin{equation}
    h_\ell^j =
\left\{
	\begin{array}{ll}
		\exp\left[-\frac{\ell(\ell+1) \theta_1^2}{16 \ln 2}\right]  & j = 1\\
		\sqrt{\exp\left[-\frac{\ell(\ell+1) \theta_j^2}{8  \ln 2}\right] - \exp\left[-\frac{\ell(\ell+1) \theta_{j-1}^2}{8 \ln 2}\right]} & 1 < j < N_{\rm bands} \\
        \sqrt{1 - \exp\left[-\frac{\ell(\ell+1) \theta_{N_{\rm bands}-1
}^2}{8 \ln 2}\right]} & j = N_{\rm bands} \\
	\end{array}
\right.
\end{equation}
where $\theta_j$ is the full width at half maximum of a Gaussian smoothing function associated to each needlet band (in radians), and $N_{\rm bands}$ is the number of needlet scales. In this work the choice for $\theta_j$ (in arcminutes) is $\{600^\prime, 300^\prime, 120^\prime, 60^\prime, 30^\prime, 15^\prime, 10^\prime\}$. The harmonic windows obtained in this way are shown in Figure \ref{fig:needlets}. Using the definition for $h_\ell^j$ we can write the needlet basis function as:
\begin{equation}
    \psi_{jk}(p) = \sqrt{\frac{4\pi}{N_j}}\sum_{\ell m} h^j_\ell Y_{\ell m}(p) Y^*_{\ell m}(\hat n_{jk})\,.
\end{equation}
In the HEALPix pixelization scheme, $N_j$ is the total number of map pixels for $j$-th needlet scale, $\hat n_{jk}$ gives the direction associated with the $k$-th pixel of the $j$-th needlet map. We can then transform the data maps $\bm d$ in terms of $\psi_{jk}$, giving coefficient maps: 
\begin{equation}
    \bm b_j(\hat n_{jk}) = \int \bm d \, \psi^*_{jk} d \Omega_p.
\end{equation}

Finally, we can compute the localized $N_{ch} \times N_{ch}$ covariance matrices from the needlet coefficient maps as:
\begin{equation}
    \bmat{ \hat C}_{jk} = \frac{1}{N_k} \sum_{k'} w_j(k, k') \bm b_j(\hat n_{jk'}) \bm b_j (\hat n_{jk'})\,,
\end{equation}
with the $w_j(k, k')$ function selecting the domain of $N_k$ pixels around the $k$-th pixel to compute the covariance at the $k$-th pixel.

\begin{figure}
    \centering
    \includegraphics[width=0.48\textwidth]{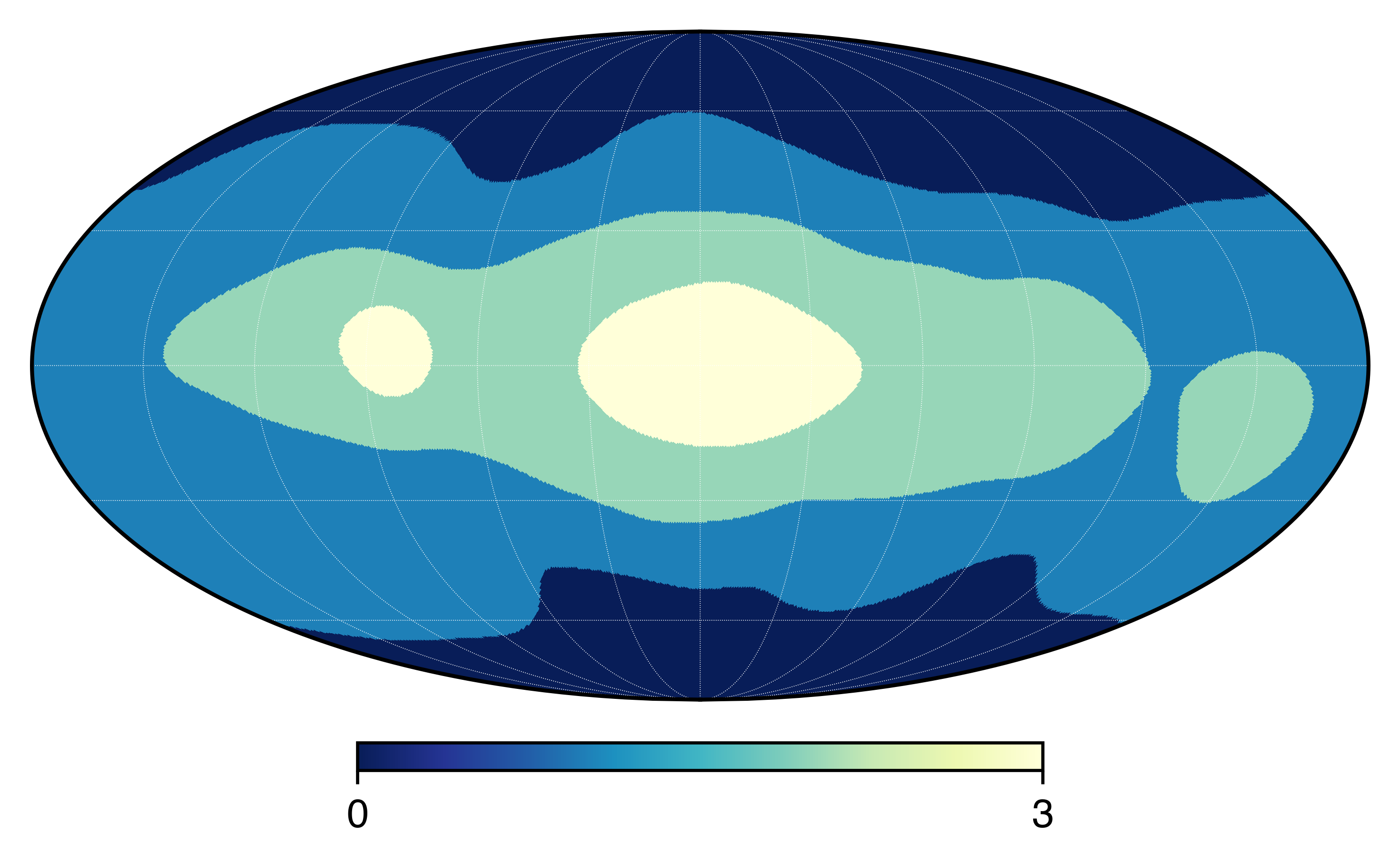}\\
    \includegraphics[width=0.48\textwidth]{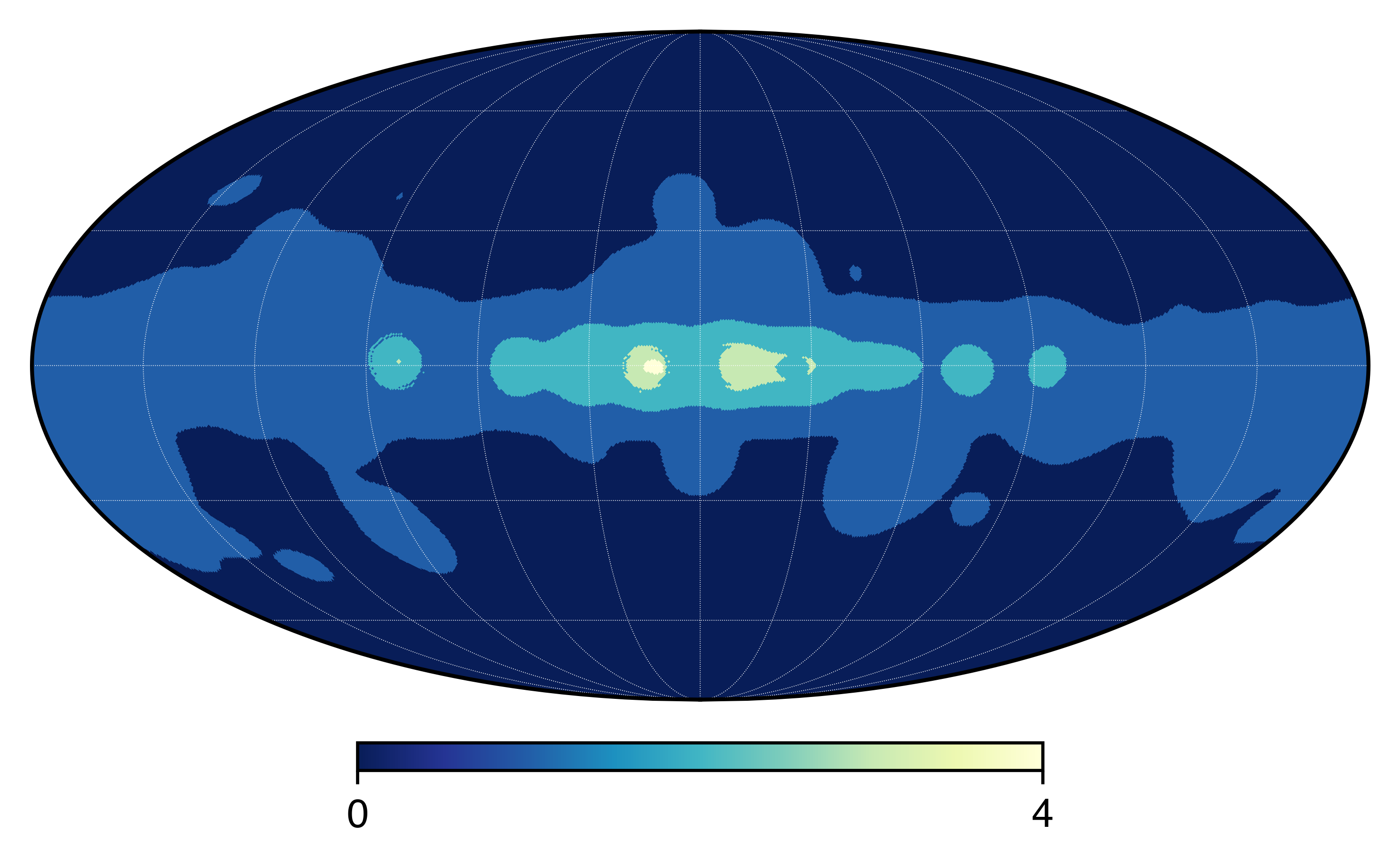}\\
    \caption{Maps of the dimension of the CO subspace as determined by GNILC 
    for the $60^\prime$ needlet band (top)
    and the $10^\prime$ needlet band (bottom). At smaller angular scales, larger parts of the high-latitude sky becomes noise-dominated (dark blue in the figure). The standard GNILC algorithm assigns those regions a zero dimensional signal subspace, which results in varying effective resolution in standard GNILC map. In our processing, in those needlet domains, we produce the CO maps using the GLS weights. With this modification to the GNILC algorithm, final xGNILC maps are produced with a uniform angular resolution.}
    \label{fig:CO_subspace}
\end{figure}
\subsubsection{Determination of the CO subspace}
The signal model introduced in Eq. \eqref{eq:signal_template} assumes that we can represent the signal in the set of input maps with a mixing matrix $\bmat A$ and templates $\bm t$. The number of emission templates required to model the signal is determined by the dimension of the `signal subspace'. Multiple dimensions are needed to describe the signal when it is bright, complex, possibly multi-component, and observed with high signal-to-noise ratio, as is the case in the Galactic plane, and in particular close to the Galactic center. One of the key features of GNILC is to determine the dimensions of the signal subspace, for every needlet band, as a function of sky position. This allows us to retain the differences between maps only where they are significant (i.e. detectable above the noise), but average down the noise contamination when the signal-to-noise ratio in regions of the pixel/harmonic space is low, and detectable astrophysical emissions nearly identical in the various observations (within uncertainties).

For the data model introduced for the Planck CO maps we can write the total covariance matrix
\begin{equation}
    \bmat{C} = \bmat{C}_s + \bmat{C}_n
\end{equation}
as the sum of the covariances of the signal and of the noise. We use the public Planck CO null maps from the Planck Legacy Archive\footnote{\url{http://pla.esac.esa.int/pla}} to get an estimate for the noise covariance $\bmat{C}_n$. We can then `whiten' the total covariance as:
\begin{equation}
    \tilde{\bmat{C}} = \bmat{C}_n^{-1/2} \bmat{C} \bmat{C}_n^{-1/2} = \bmat{C}_n^{-1/2} \bmat{C}_s \bmat{C}_n^{-1/2} + \bmat{I}.
    \label{eq:white_cov}
\end{equation}
In general, we can then obtain a eigenvalue decomposition of the $\tilde{\bmat C}$ as:
\begin{equation}
    \tilde{\bmat C} = \bmat U \bmat D \bmat U^T = \bmat U_s \bmat D_s \bmat U^T_s + \bmat U_n\bmat U_n^T,
    \label{eq:cov_eigen}
\end{equation}
where we use Eq. \eqref{eq:white_cov}. Here $\bmat U_s$ is a $N_{ch} \times N_s$ matrix with $N_s$ columns representing  eigenvectors for the eigenvalues associated with the signal subspace. As discussed in \cite{planck2016-XLVIII}, we can obtain the number of dimensions, $N_s$, of the signal subspace from the eigenvalues, $\bmat D = {\rm diag}(\lambda_1,\dots, \lambda_\alpha,\dots, \lambda_{N_{ch}})$, by minimizing the Akaike Information Criterion
\begin{equation}
    \min_{\alpha \in [1, N_{ch}]} \left[ 2\alpha + \sum_{k=\alpha+1}^{N_{ch}} (\lambda_k -\log \lambda_k -1)\right].
\end{equation}
This gives an estimate of the CO subspace dimension, $N_s$, as a function of sky position for every needlet scale. In Fig. \ref{fig:CO_subspace} we show the GNILC estimated CO subspace dimension for the $60^\prime$ and $10^\prime$ needlet scales. It is clear from the figure that for needlet scales covering smaller scales, the signal-to-noise ratio gets poorer further away from the Galactic plane. Consequently, the traditional GNILC method sets the CO signal subspace dimension to be zero in these noise-dominated regions of needlet space.

The signal covariance matrix can be written as:
\begin{equation}
    \bmat C_s = \bmat A \langle \bm t \bm t^T \rangle \bmat A^T.
\end{equation}
Using Eqs. \eqref{eq:white_cov} and \eqref{eq:cov_eigen} we can also write the signal covariance matrix as:
\begin{equation}
    \bmat C_s = \bmat C_n^{1/2} \bmat U_s (\bmat D_s - \bmat I)\bmat U_s^T \bmat C_n^{1/2}.
\end{equation}
We then estimate the mixing matrix as $\bmat {\hat A} = \bmat C_n^{1/2} \bmat U_s$. 

\subsubsection{Multidimensional ILC and GLS}
We produce cleaned maps, $\bm{\hat s}$, as a combination of $N_{ch}$ input maps, $\bm d$, with a set of optimization weights from our extended GNILC pipeline, $\bmat W_{\rm xGNILC}$, given by
\begin{equation}
    \bm{\hat s} = \bmat W_{\rm xGNILC} \bm d.
\end{equation}
\cite{Remazeilles:2011} have shown that the multidimensional ILC weights are
\begin{equation}
    \bmat W_{\rm ILC} = \bmat {\hat A} (\bmat {\hat A}^T \bmat C^{-1} \bmat {\hat A})^{-1} \bmat {\hat A}^T \bmat C^{-1}.
\end{equation}
We use the mixing matrix estimated in the previous step to compute the multidimensional ($N_s$-dimensional) ILC weights. 

In regions of the map that have poor signal-to-noise, the GNILC algorithm estimates the signal subspace dimension to be zero (Fig.~\ref{fig:CO_subspace}). This typically happens for regions at high Galactic latitude, for needlet bandpasses that cover the smaller scales, where the CO line emission is very faint and the noise is comparatively high. As a consequence, the traditional GNILC algorithm would  assign zero weights for all and throw away all information in that needlet domain. As mentioned earlier, this results in two issues: 1 - variation of map resolution from location to location on the sky, and 2 - loss of faint signal. 
To avoid this drawback, we instead perform an inverse noise weighted combination by inverting the system $\bm d = \bar{\bm a} t + \bm n$, where $\bar{\bm {a}}$ is the $N_{ch}$-dimensional `mixing vector' determined in the calibration step (Section \ref{sec:calibration}), and $t$ is a single CO template. This inverse noise weighted solution is obtained through the Generalized Least Squares (GLS) solution:
\begin{equation}
    \bmat W_{\rm GLS} = \bar{\bm a} (\bar{\bm a}^T \bmat C_n^{-1} \bar{\bm a})^{-1} \bar{\bm a}^T \bmat C_n^{-1}.
    \label{eq:gls}
\end{equation}
In summary, we can write the xGNILC weights as:
\begin{equation}
    \bmat W_{\rm xGNILC} =
\left\{
	\begin{array}{ll}
		\bmat W_{\rm ILC}  & N_s > 0\\
		\bmat W_{\rm GLS} & N_s = 0 \\
	\end{array}
\right .
\end{equation}
\subsubsection{Recombination into processed maps}
We obtain re-processed maps from the GNILC cleaned needlet coefficient maps 
by inverse needlet transform, 
at HEALPix $N_{\rm side}$ of 1024 and at $10^\prime$ resolution. We finally replace the regions of very bright CO that were masked/saturated as described in Section \ref{sec:preproc} with data in the same patch from the corresponding original input Planck CO data product. 

\begin{figure*}
    \centering
    \includegraphics[width=0.47\textwidth]{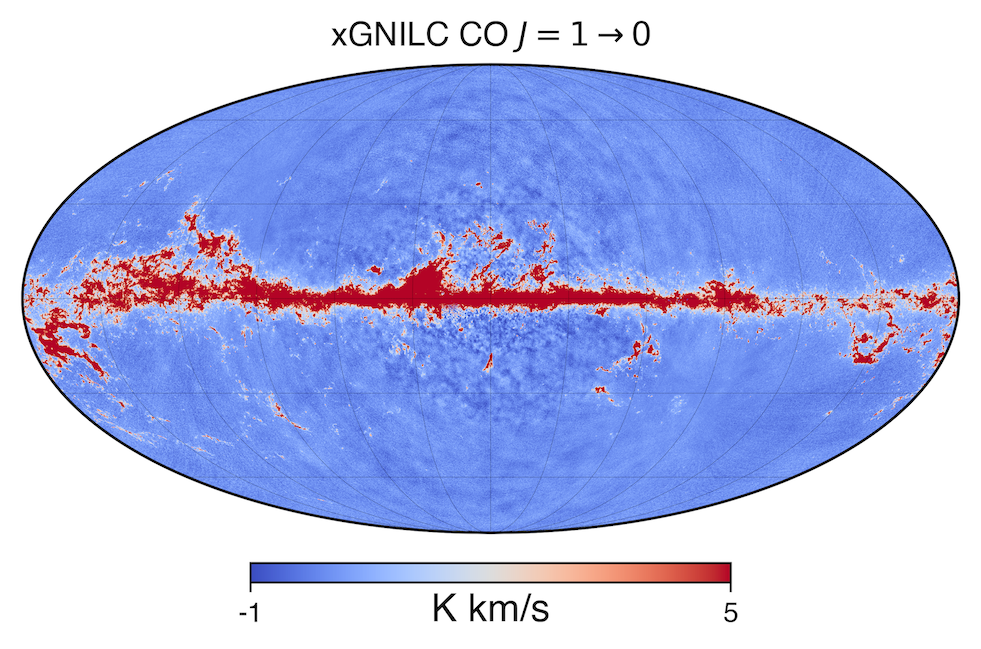}
    \includegraphics[width=0.47\textwidth]{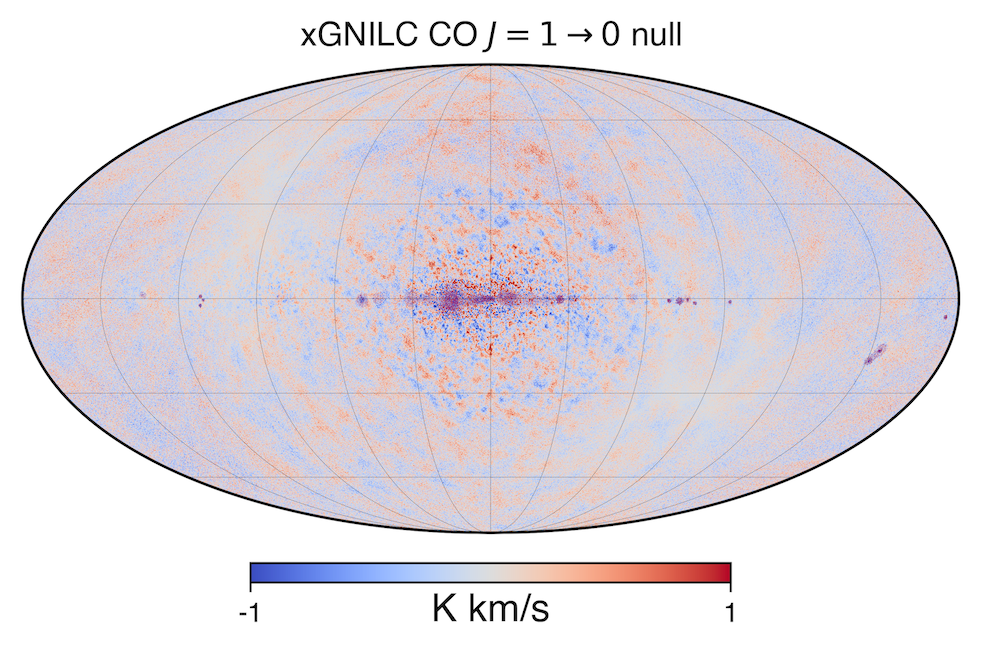}\\
    \includegraphics[width=0.47\textwidth]{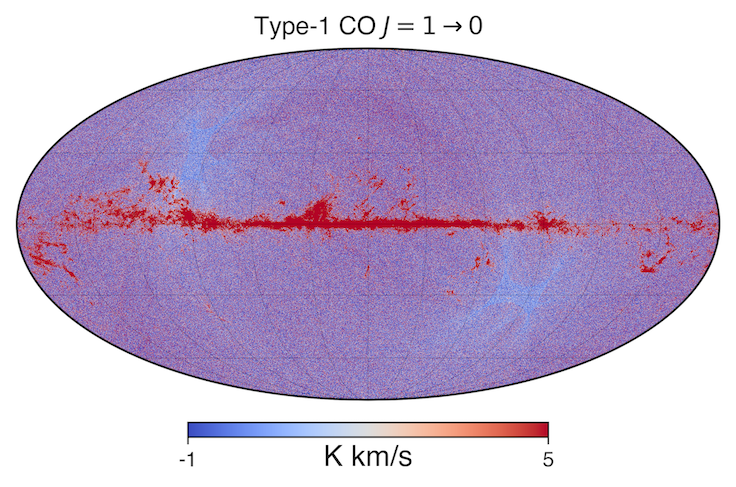}
    \includegraphics[width=0.47\textwidth]{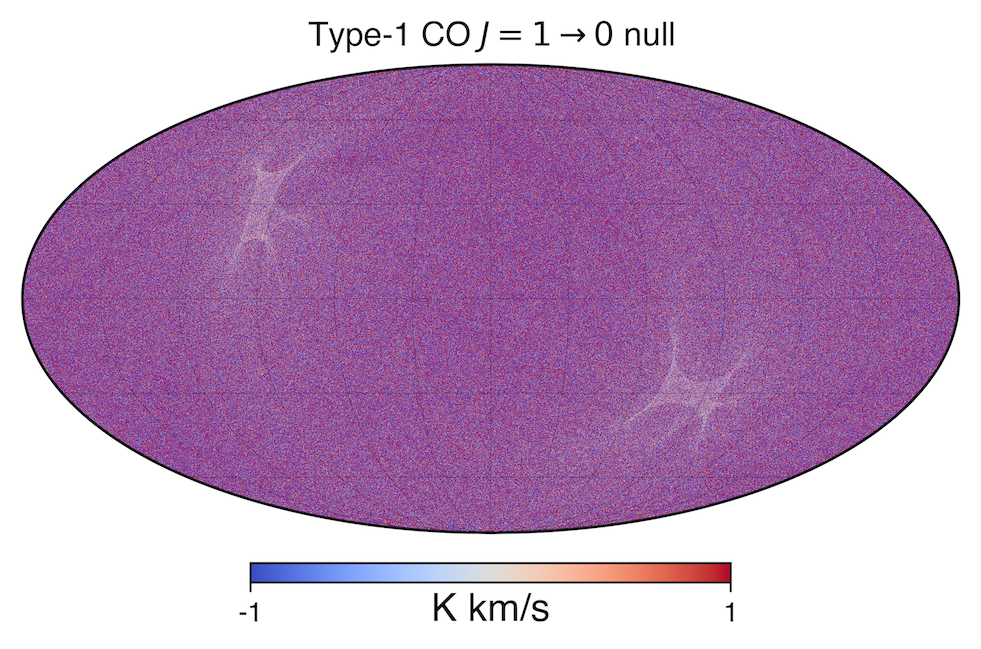}
    \caption{Top: the xGNILC CO \coj\ map, at output resolution of $10^\prime$, and the corresponding jackknife noise map. Bottom: Similar maps for the Planck \typeone\ CO \coj\ data product, and the jackknife noise of the Planck \typeone\ CO \coj\ map, both smoothed to $10^\prime$ resolution. 
    }
    \label{fig:co10_data-prod}
\end{figure*}

\subsection{Postprocessing}
\label{sec:postprocessing}
The postprocessing step is only relevant for the CO \cojjj\ data product from our pipeline. As discussed in Sec. \ref{sec:limitations_of_existing_maps}, the CO \cojjj\ has clearly visible systematic artifacts around the Galactic plane. 
These artifacts are not removed by the xGNILC pipeline.\footnote{They should not. For GNILC, these patterns are above the noise threshold, and identified as signal that is only present in the \cojjj\ maps.} 
In the postprocessing stage we reduce contamination by those residual systematics by filtering. In practice, we filter a difference map
\begin{equation}
    \Delta = d_{\rm xGNILC}^{J=3\to 2} - \left(\bar a_{J=3\to 2} / \bar a_{J=2\to 1}\right) d_{\rm xGNILC}^{J=2\to 1},
\end{equation}
i.e. a \cojjj\ map in which the CO signal is reduced  
by subtracting the \cojj\ signal rescaled with the ratio of the global calibration factors. 
As the scaling relation is only approximate, we find several CO-associated features in the residual map, due to the CO signal not canceling out. We threshold that difference map and mask all pixel where the absolute value of the residual is above 0.8 ${\rm K_{RJ}\, km \, s}^{-1}$. 
We then median filter the masked residual map in superpixels at $N_{\rm side}=16$, and smooth that median-filtered map with a $90^\prime$ Gaussian beam. The resulting map captures the large scale features of the residuals shown in Fig. \ref{fig:type1-J3-2}. We subtract this median-filtered  masked residual map from the xGNILC CO \cojjj\ map to produce our released data product for \cojjj. 

\subsection{Uncertainties}

The error in the CO maps comprises a statistical uncertainty from the measurement noise, and a systematic uncertainty due to the choice of the calibration factor that is used as a prior.
Unfortunately, it is not possible to estimate the statistical uncertainty for the xGNILC product. The input Planck \typeone\ and \typetwo\ maps come with their individual uncertainty maps, but \typeone\ and \typetwo\ maps are not independent and they have significant cross-covariances. This information is unavailable, making it nearly impossible to ensure correct statistics of any simulated noise maps. We obtain maps of the systematic uncertainty which is shown in Fig. \ref{fig:xgnilc-sys}.\footnote{On top of this, any calibration uncertainty in the original maps is still present in our final products.}

\subsubsection{Systematic uncertainties}
The systematic uncertainty captures the impact of calibration uncertainties that enter the pipeline through the choice of a prior on the mixing vector $\bm{\bar a}$ (but not any calibration uncertainties already impacting the original Planck data products). For the xGNILC CO \cojjj\ map the calibration uncertainty also impacts the postprocessing step as we use the global scaling ratio to compute the residual map. We use the individual fit over 30 sub-regions (with offset), as discussed in Sec. \ref{sec:calibration}, to construct a covariance matrix for the scaling parameter $a_\alpha$. This covariance matrix with the mean values for $a_\alpha$, given in Tab. \ref{tab:calibration}, is used to define a multivariate Gaussian distribution for the mixing vector. 

\begin{figure*}
    \centering
    \includegraphics[width=0.47\textwidth]{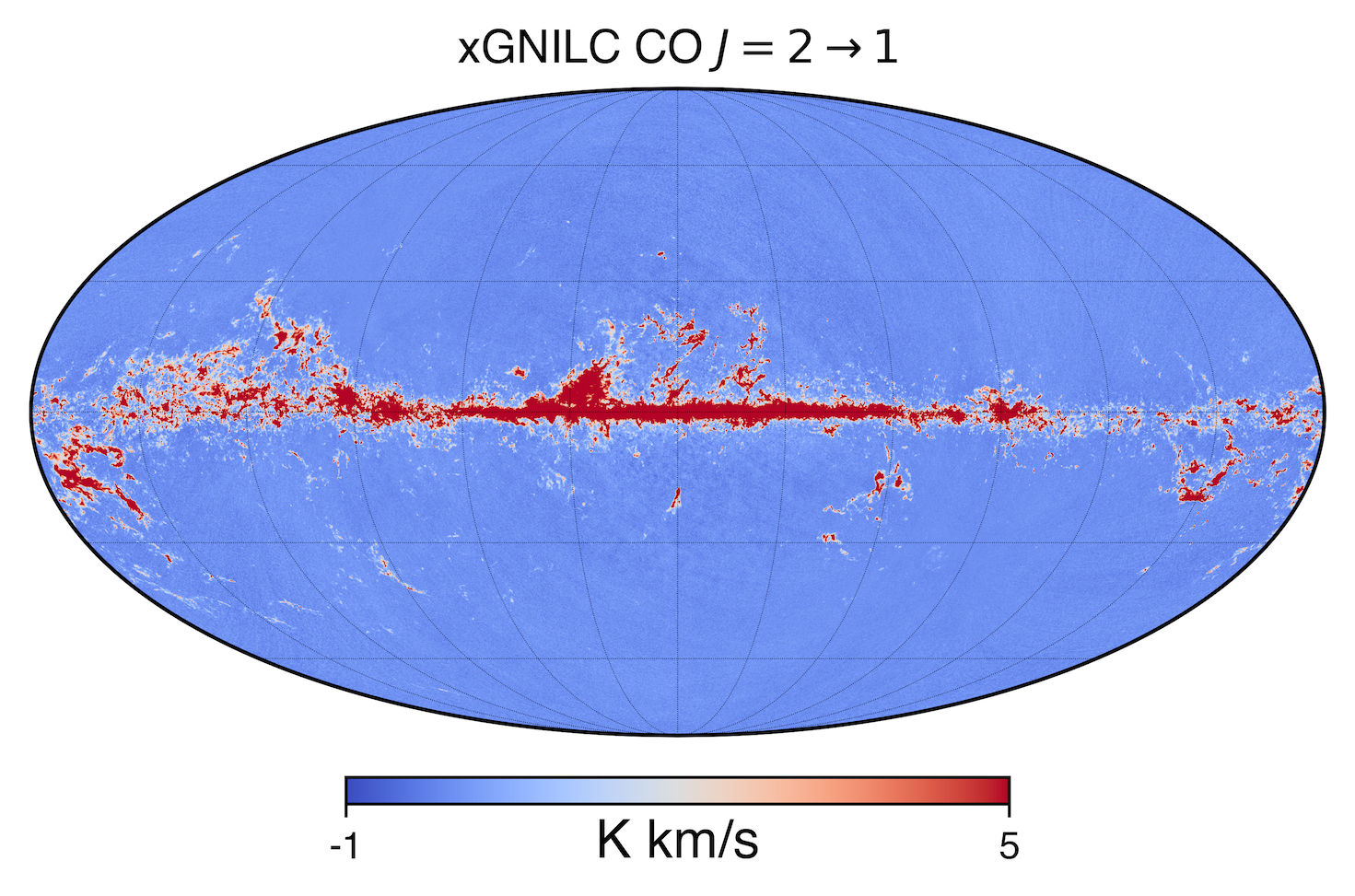}
    \includegraphics[width=0.47\textwidth]{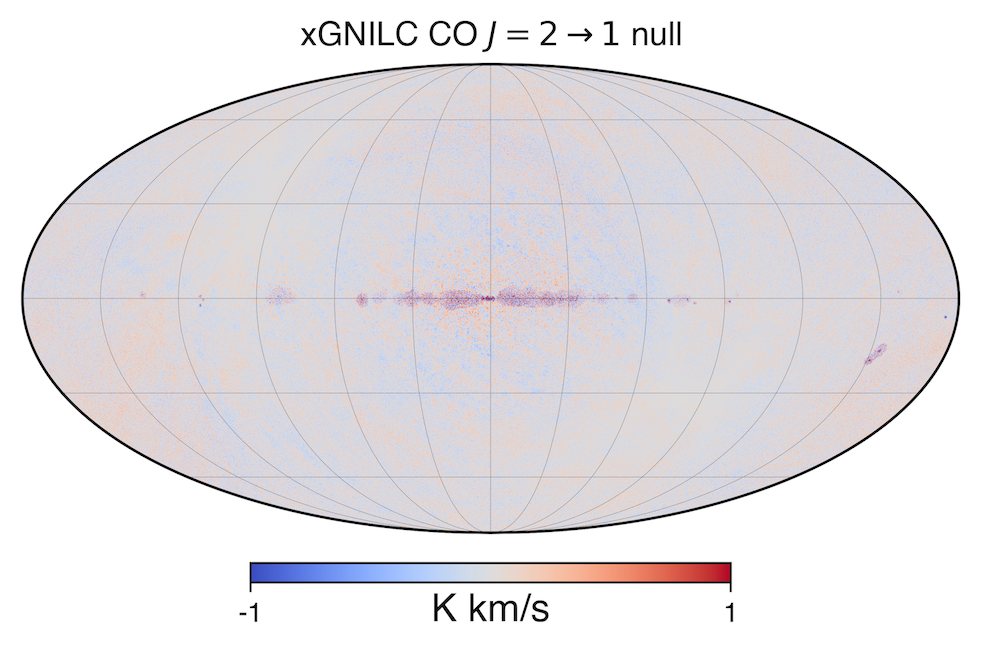}\\
    \includegraphics[width=0.47\textwidth]{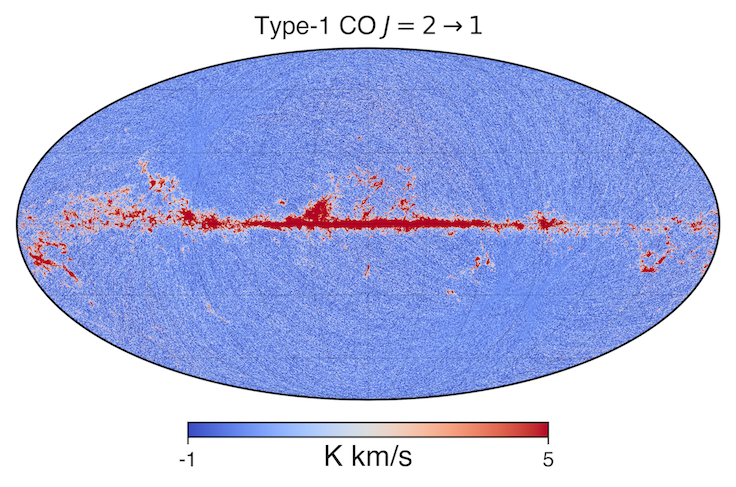}
    \includegraphics[width=0.47\textwidth]{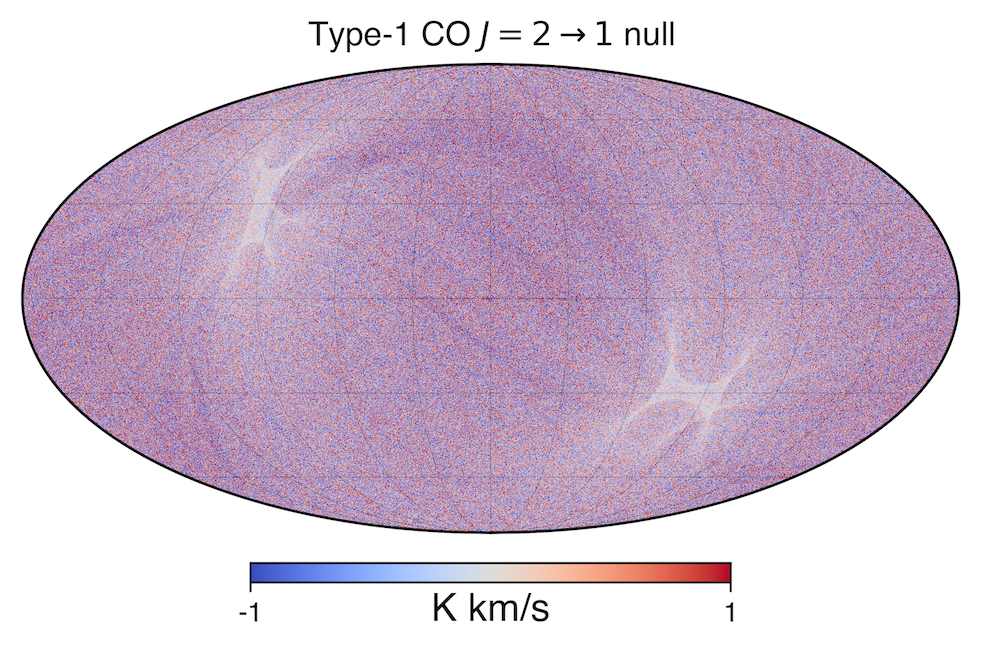}
    \caption{The xGNILC CO \cojj\ map, at output resolution of $10^\prime$, and the jackknife noise map for the xGNILC CO \cojj\ map are shown in the top row. In the bottom row, the Planck \typeone\ CO \cojj\ map and the jackknife noise of the Planck \typeone\ CO \cojj\ map both smoothed to $10^\prime$ resolution.}
    \label{fig:co21_data-prod}
\end{figure*}

Systematic uncertainties are then obtained by running the xGNILC pipeline 
with 200 different multivariate Gaussian realizations of the prior used for $\bm{\bar a}$. For the CO \cojjj\ map we perform the additional step of postprocessing the maps to correct for the residual systematics, with different value for the ratio of scaling factors. We compute the standard deviation of the 200 different output maps we obtained in the process. For the CO \coj\ and \cojj\ maps the systematic uncertainty is only nonzero in the regions away from the Galactic plane where the xGNILC pipeline used the GLS weights. In case of the CO \cojjj\ maps the systematic uncertainty of the postprocessing step is larger than the GLS step, and we see more systematic uncertainty in the Galactic plane. However, for all three maps, the systematic uncertainties are still negligible compared to the statistical uncertainty.

\subsubsection{Contamination}

Our basic xGNILC maps are obtained by combining \typeone\ and \typetwo\ maps. This generates the results with the lowest noise contamination, but the possibility of contamination by other foregrounds originating from the \typetwo\ maps cannot be ruled-out. 

We hence repeat the whole process using only \typeone\ maps (i.e. three input maps instead of five). We then compute the difference between xGNILC products obtained in the two cases and we flag for possible significant contamination, for each of the CO lines, map pixels where all of the following three conditions are met:
\begin{enumerate}
    \item a discrepancy between the two data products (xGNILC with or without \typetwo\ maps) is detected locally at more than $4\sigma$ in the xGNILC difference map;
    \item the difference is larger than 15\% of the local xGNILC CO emission (i.e. we tolerate discrepancies that do not change the CO emission by more than 15\%);
    \item the CO signal itself in the final xGNILC CO map is detected locally at more than $4\sigma$ (i.e. we don't flag regions where there is very low signal, barely detectable in our final maps).
\end{enumerate}
For each of the three CO lines, a confidence mask is generated by excluding a circle of radius $30^\prime$ around each pixel flagged in this way. We concatenate these three masks into a single confidence mask.

\section{Data products}
\label{sec:data-products}

The xGNILC CO data products contain, for each CO line, the following set of maps:
\begin{itemize}
    \item xGNILC CO map at $N_{\rm side}=1024$. 
    \item xGNILC CO null map at $N_{\rm side}=1024$ obtained by projecting the Planck null maps with xGNILC weights.
    \item Systematic uncertainty maps for the xGNILC data products.
    \item All the masks used in the analysis and confidence mask for the data product. 
    \item The beam transfer function of xGNILC data products.
\end{itemize}

The xGNILC CO \coj\ map and projected null map are shown in Fig. \ref{fig:co10_data-prod} along with corresponding figures showing the Planck \typeone\ \coj\ data product. 
Both are for an angular resolution of $10^\prime$. 

\begin{figure*}
    \centering
    \includegraphics[width=0.47\textwidth]{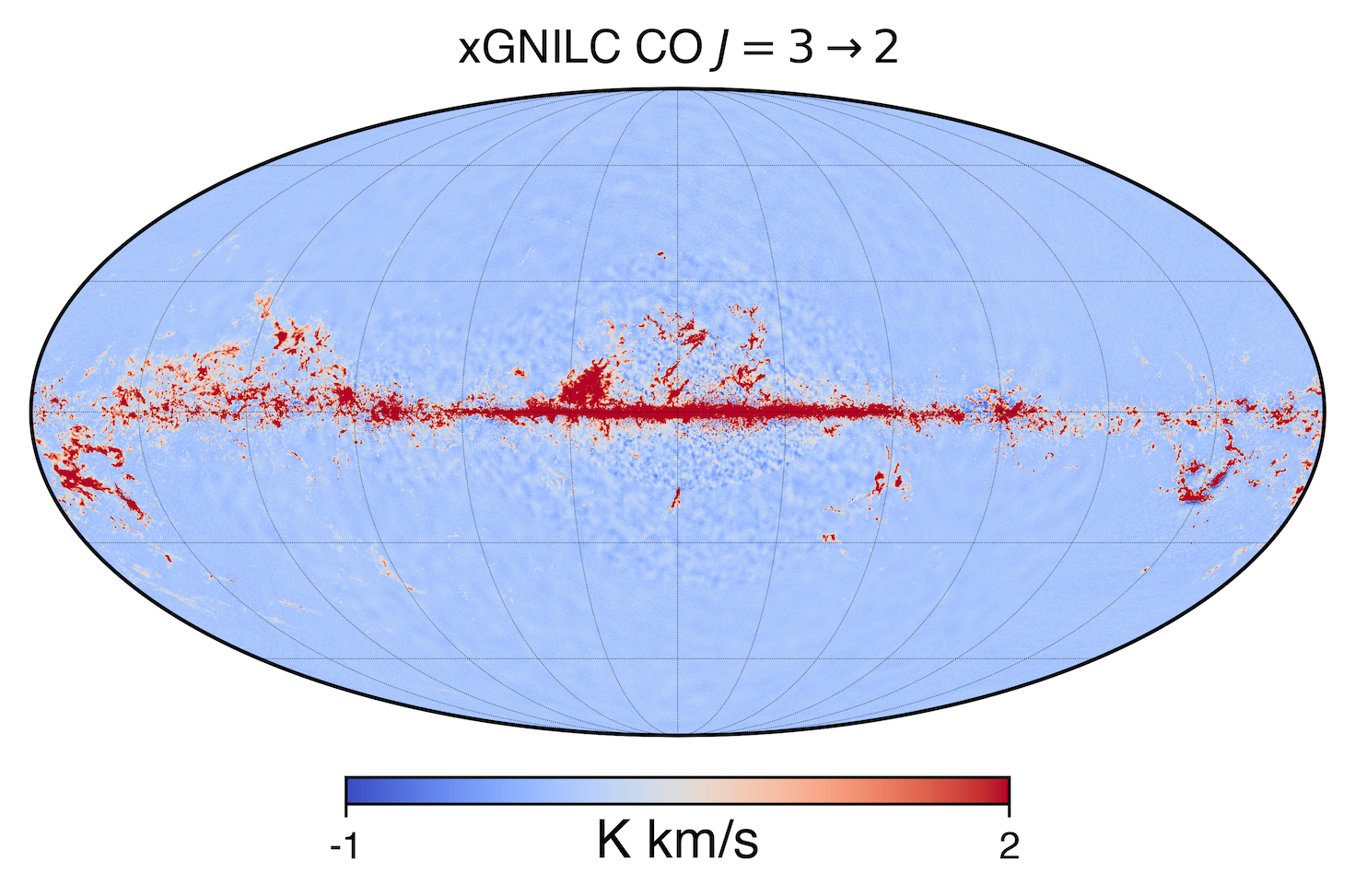}
    \includegraphics[width=0.47\textwidth]{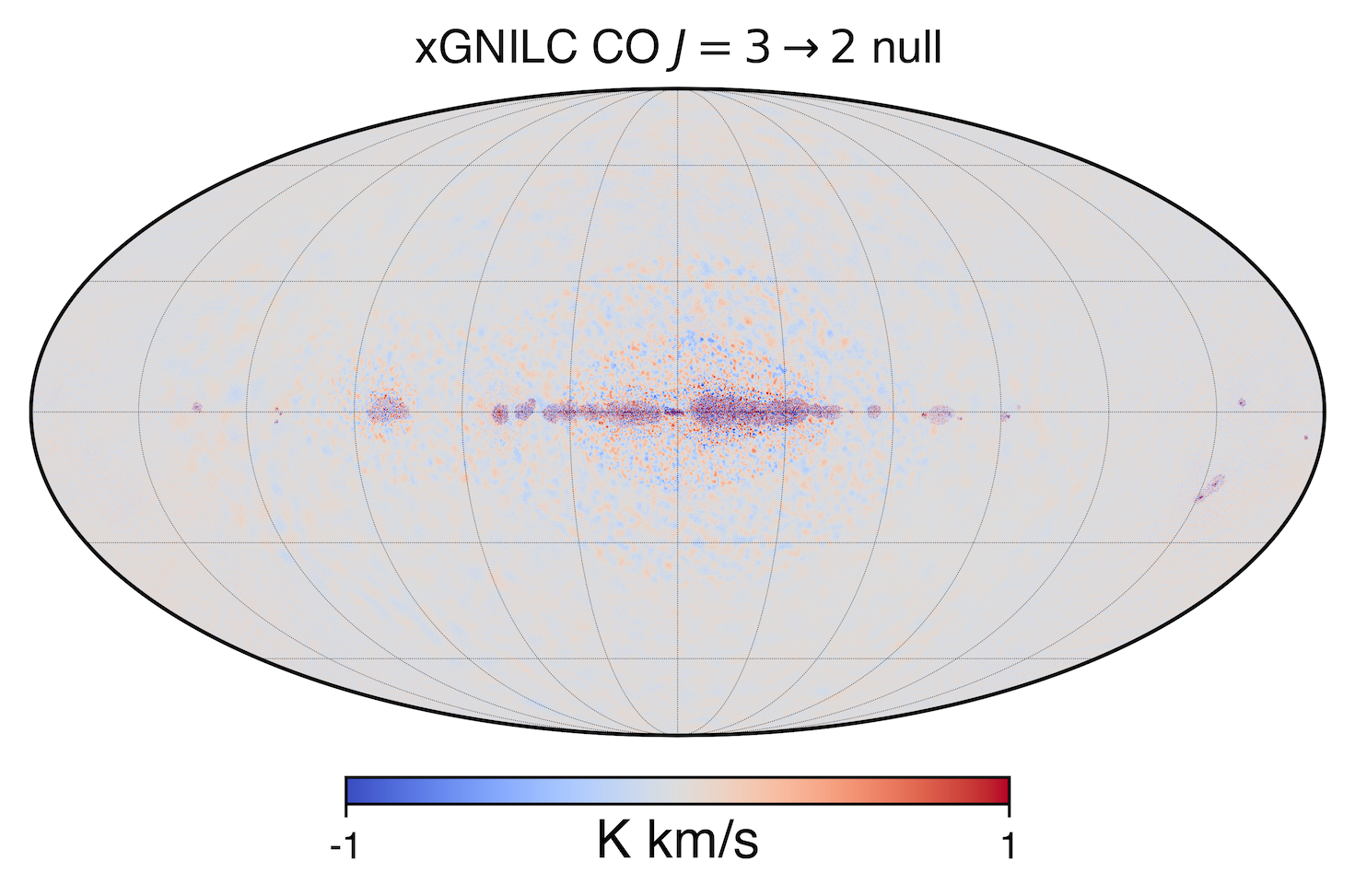}\\
    \includegraphics[width=0.47\textwidth]{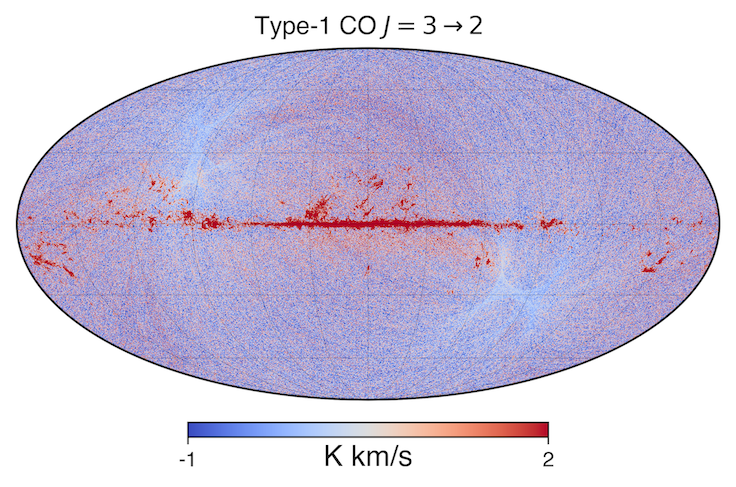}
    \includegraphics[width=0.47\textwidth]{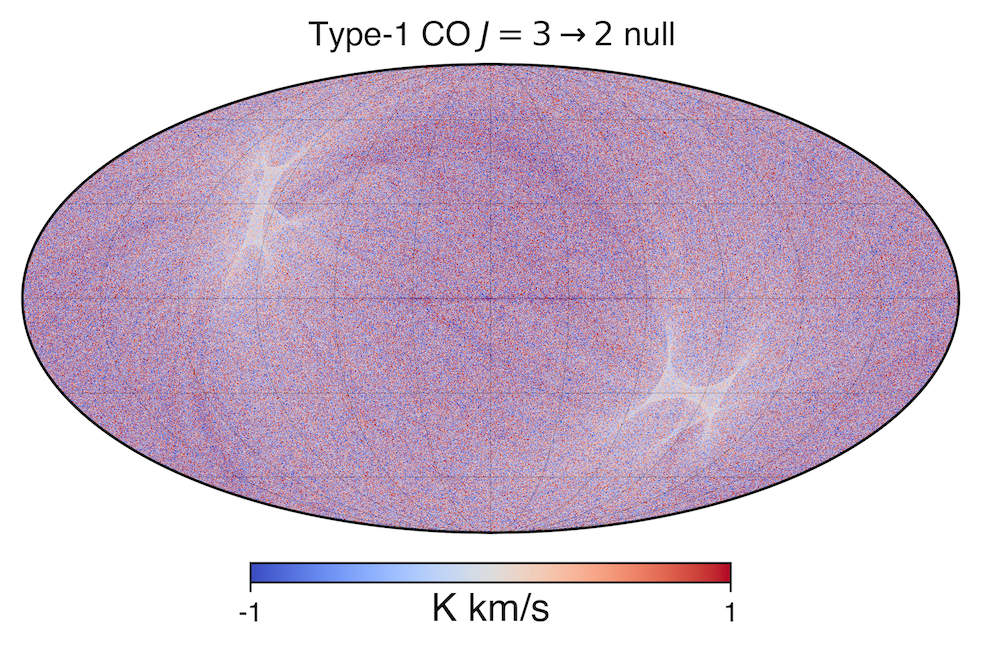}
    \caption{The xGNILC CO \cojjj\ map, at output resolution of $10^\prime$, and the jakcknife noise for the xGNILC CO \cojjj\ map are show in the top row. In the bottom row, the Planck \typeone\ CO \cojjj\ map and the corresponding jackknife noise map smoothed to $10^\prime$ resolution.}
    \label{fig:co32_data-prod}
\end{figure*}

A visual inspection of the maps (CO map and null) show a spectacular improvement in signal-to-noise of the CO \coj\ line emission in the new xGNILC data product. We note that the structure of the xGNILC null map is more complex, with clear distinction between the behavior near the Galactic center, most regions in the Galactic plane, and outside the Galactic CO emission region. The noise behavior is largely driven by the number of foreground subspace dimensions estimated by GNILC in the last needlet band, except for the masked bright sources such as the Galactic center, for which noise properties are identical to those of the corresponding \typeone\ input map. 
The immediate region near the Galactic center is more noisy than the rest of the sky as GNILC tends to find higher foreground subspace dimensions in regions near the Galactic center. 

\begin{figure*}
    \centering
    \includegraphics[width=0.33\textwidth]{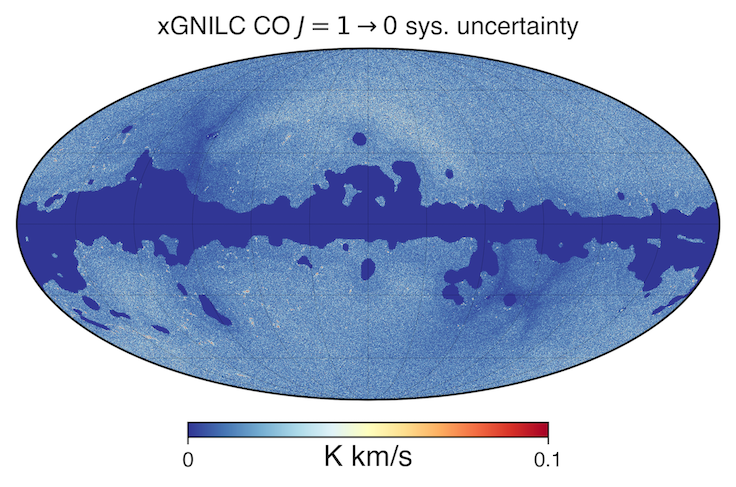}
    \includegraphics[width=0.33\textwidth]{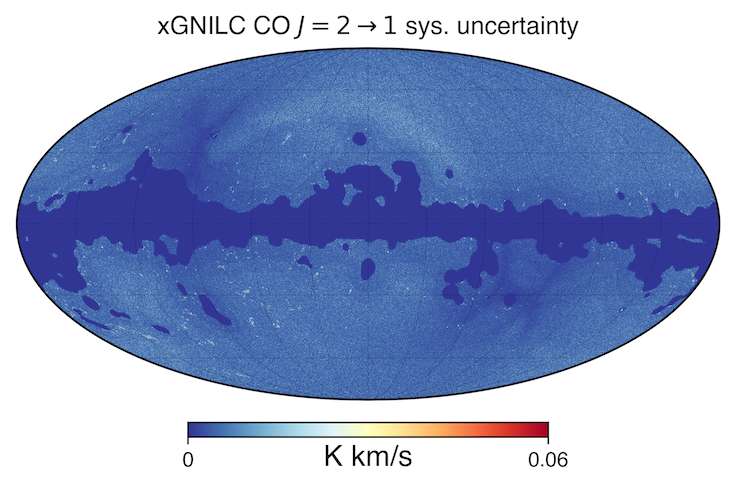}
    \includegraphics[width=0.33\textwidth]{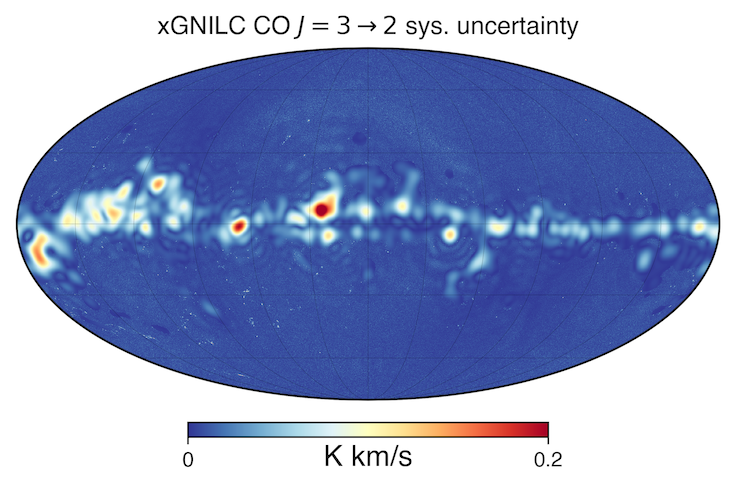}
    \caption{Maps of systematic uncertainty in the xGNILC data products arising from uncertainty in the priors used for the CO scaling parameters. The uncertainty in the CO \coj\ and \cojj\ maps are only in the region where we do a GLS and so it does not contribute to any uncertainty in the Galactic plane. For the CO \cojjj\ map, the dominant systematics uncertainty come from the postprocessing stage.}
    \label{fig:xgnilc-sys}
\end{figure*}


The xGNILC CO \cojj\ map and projected null map are shown in Fig. \ref{fig:co21_data-prod}, with the corresponding \typeone\ map and null are reconvolved with the xGNILC beam transfer function. 
The \typeone\ CO \cojj\ map has better signal-to-noise than other Planck \typeone\ maps. Again,  the xGNILC CO \cojj\ map has visibly better signal-to-noise level than the original \typeone\ CO \cojj\ map. The discussion above for noise in the different regions of the sky for xGNILC CO \coj\ applies for the CO \cojj\ map as well.

In Fig. \ref{fig:co32_data-prod} we compare the xGNILC CO \cojjj\ map and projected null map with those of the Planck \typeone\ CO \cojjj\ map, smoothed  for a comparison at $10^\prime$ resolution of the xGNILC map. 
The xGNILC \cojjj\ map is visibly less noisy, and the systematics around the Galactic ridge have been removed. The xGNILC noise is significantly lower than the uncertainty of the corresponding \typeone\ map. 

\begin{figure*}
    \centering
    \includegraphics[width=0.8\textwidth]{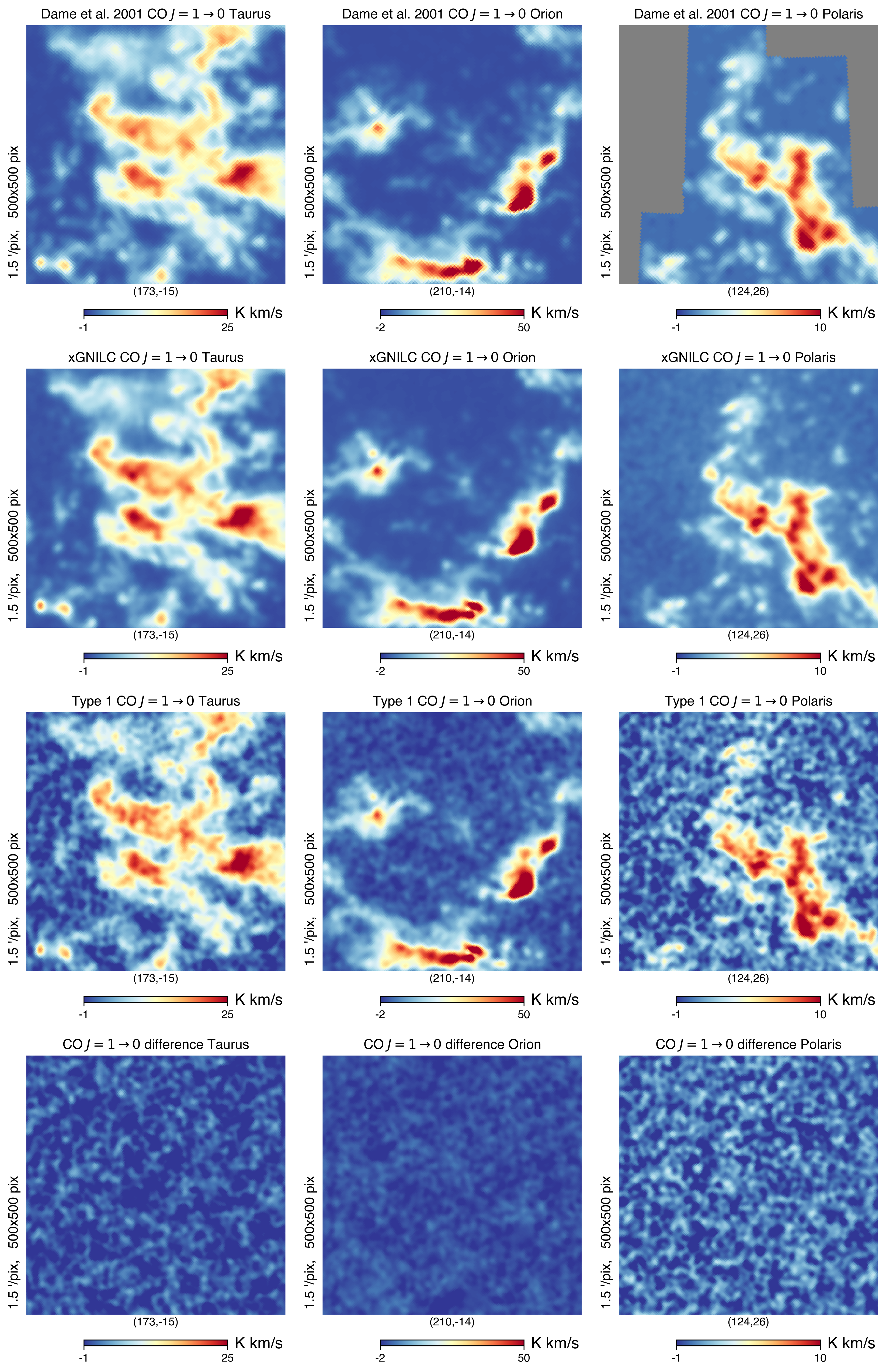}
    \caption{Images $(12.5^\circ \times 12.5^\circ)$ of important CO emission regions for CO \coj\ : Taurus (left column), Orion (middle column), and Polaris (right column), at 20' resolution. The first row shows the \citet{dame2001} map (unobserved pixels are shown in gray), second row shows our xGNILC map, third row shows the Planck \typeone\ map and in the last row we show the difference map (Planck \typeone\ $-$ xGNILC). We have smoothed these maps to 20' to suppress the noise in the Planck \typeone\ map. Note the spectacular consistency 
    between xGNILC maps and Dame maps.}
    \label{fig:co10_region_comparison}
\end{figure*}

\begin{figure*}
    \centering
    \includegraphics[width=0.8\textwidth]{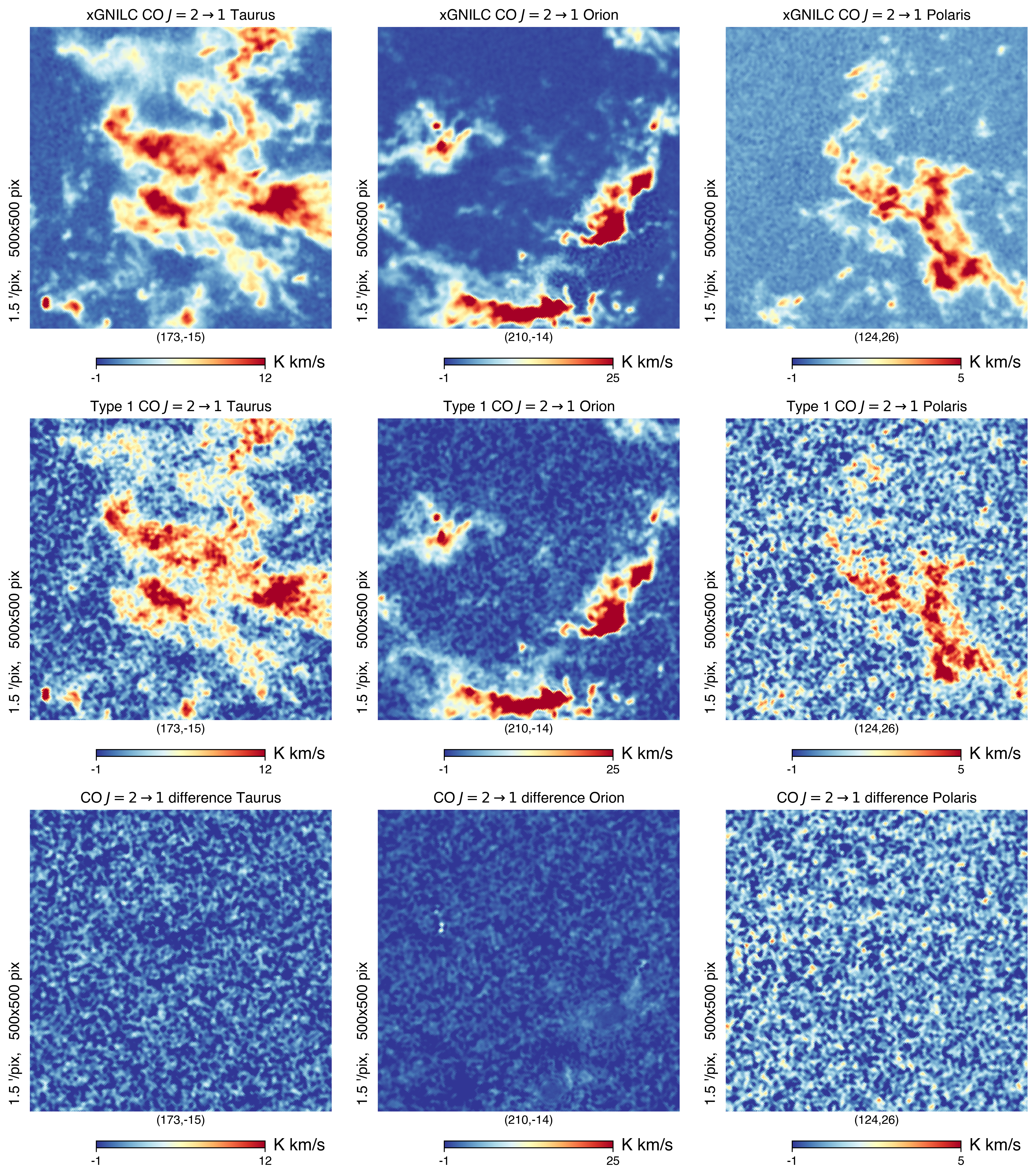}
    \caption{Images $(12.5^\circ \times 12.5^\circ)$ of important CO emission regions for CO \cojj\ : Taurus (left column), Orion (middle column), and Polaris (right column), at $10^\prime$ resolution. The first row shows our xGNILC map, the second row shows the Planck \typeone\ map, and the difference map (Planck \typeone\ $-$ xGNILC) is shown in the third row.}
    \label{fig:co21_region_comparison}
\end{figure*}


In Fig. \ref{fig:xgnilc-sys} we show the estimate of systematic uncertainty due to the uncertainty in the choice of priors on the mixing vector. For the xGNILC CO \coj\ and \cojj\ data products, the systematic uncertainty is only non-zero in the region outside the Galactic plane where we do a GLS. For the CO \cojjj\ map, there is more contamination in the Galactic plane because of the postprocessing stage which also assumes this prior. The systematic uncertainty is sub-dominant for all three xGNILC data products.

In the preprocessing step of the pipeline we have masked most of the thermal SZ clusters present in the \typetwo\ maps (which would contaminate the xGNILC data products otherwise). However, all input maps also have point source contamination which are not removed and propagates to xGNILC data products. Additionally, there are few regions of bright extra-galactic CO emission from our local neighborhood, like the Large Magellenic Cloud, Small Magellenic Cloud, M81 etc. Given the different use cases for the data products, we leave it to the end user to mitigate these contaminations as per requirement.

\section{Validation and discussion}
\label{sec:discussion}

For validation, we compare the xGNILC maps with the existing Planck CO data products. Planck products validation has been extensively discussed in \cite{planck2013-p03a}, and we make no attempt to re-validate the Planck CO maps as compared to other external data sets. 

\begin{figure*}
    \centering
    \includegraphics[width=0.8\textwidth]{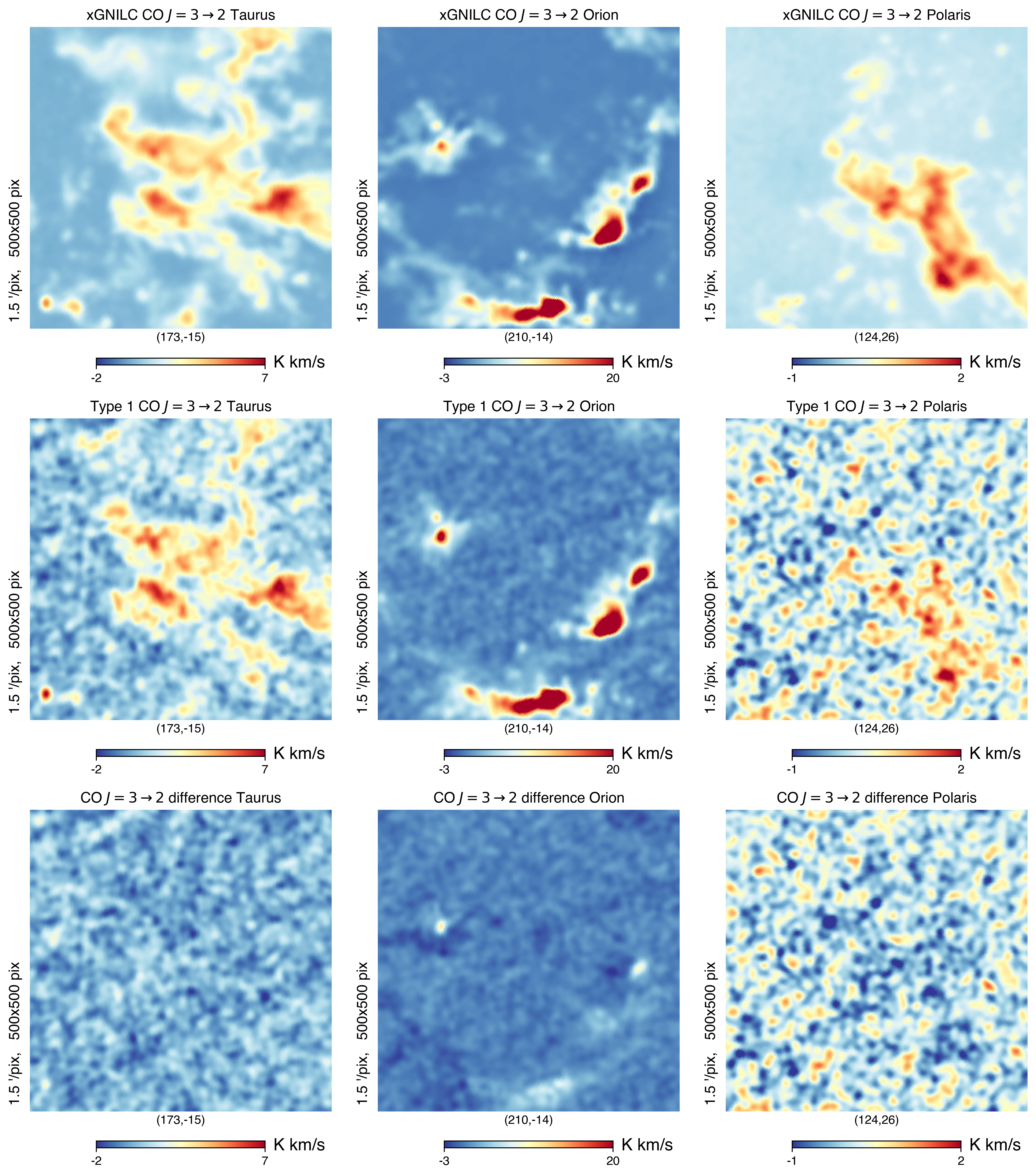}
    \caption{Images $(12.5^\circ \times 12.5^\circ)$ of important CO emission regions for CO \cojjj\ : Taurus (left column), Orion (middle column), and Polaris (right column), at 20' resolution. The first row shows our xGNILC map, the second row shows the Planck \typeone\ map, and the third row shows the difference map (Planck \typeone\ $-$ xGNILC). We have additional smoothing for this comparison as the signal-to-noise level of the Planck map is low. }
    \label{fig:co32_region_comparison}
\end{figure*}

\subsection{Map-level validation}
In \cite{planck2013-p03a} three specific regions of CO emission where studied for validation: Taurus $(l=173^\circ, b=-15^\circ)$, Orion $(l=210^\circ, b=-14^\circ)$, and Polaris $(l=124^\circ, b=26^\circ)$. In Fig. \ref{fig:co10_region_comparison} - \ref{fig:co32_region_comparison} we compare the xGNILC maps and Planck \typeone\ maps for these three regions. We also show the difference map (Planck \typeone\ - xGNILC) for these regions. For the CO \coj\ map comparison we also show the Dame CO \coj\ map. The CO \coj\ and \cojjj\ line emissions map comparisons are at a $20^\prime$ resolution to suppress the noise in the Planck \typeone\ maps. 

The visual comparison for CO \coj\ maps shows excellent agreement between the Dame and xGNILC maps, with a bit of excess in the xGNILC maps, as expected given the relative calibration coefficients, and consistent with a contribution from CO isotopologues. The Planck \typeone\ map remains noisy even after smoothing to $20^\prime$, but shows features consistent with the xGNILC map. Finally, the difference of the Planck \typeone\ and xGNILC map shows no difference that is visually different from noise. 

In Fig. \ref{fig:co21_region_comparison} we present a similar comparison for the CO \cojj\ maps. The comparison for this line, for which the signal-to-noise ratio is higher than for the other two, is done at the $10^\prime$  resolution of the xGNILC map. We see good agreement between the features seen in the xGNILC and the Planck \typeone\ maps. The difference map for CO \cojj\ shows some small ringing features that are seen in the xGNILC map in the Orion region, but these features are small in comparison to the brightness of the CO signal in this region. The difference maps in the other two regions are consistent with noise.

\begin{figure*}
    \centering
    \includegraphics[width=0.8\textwidth]{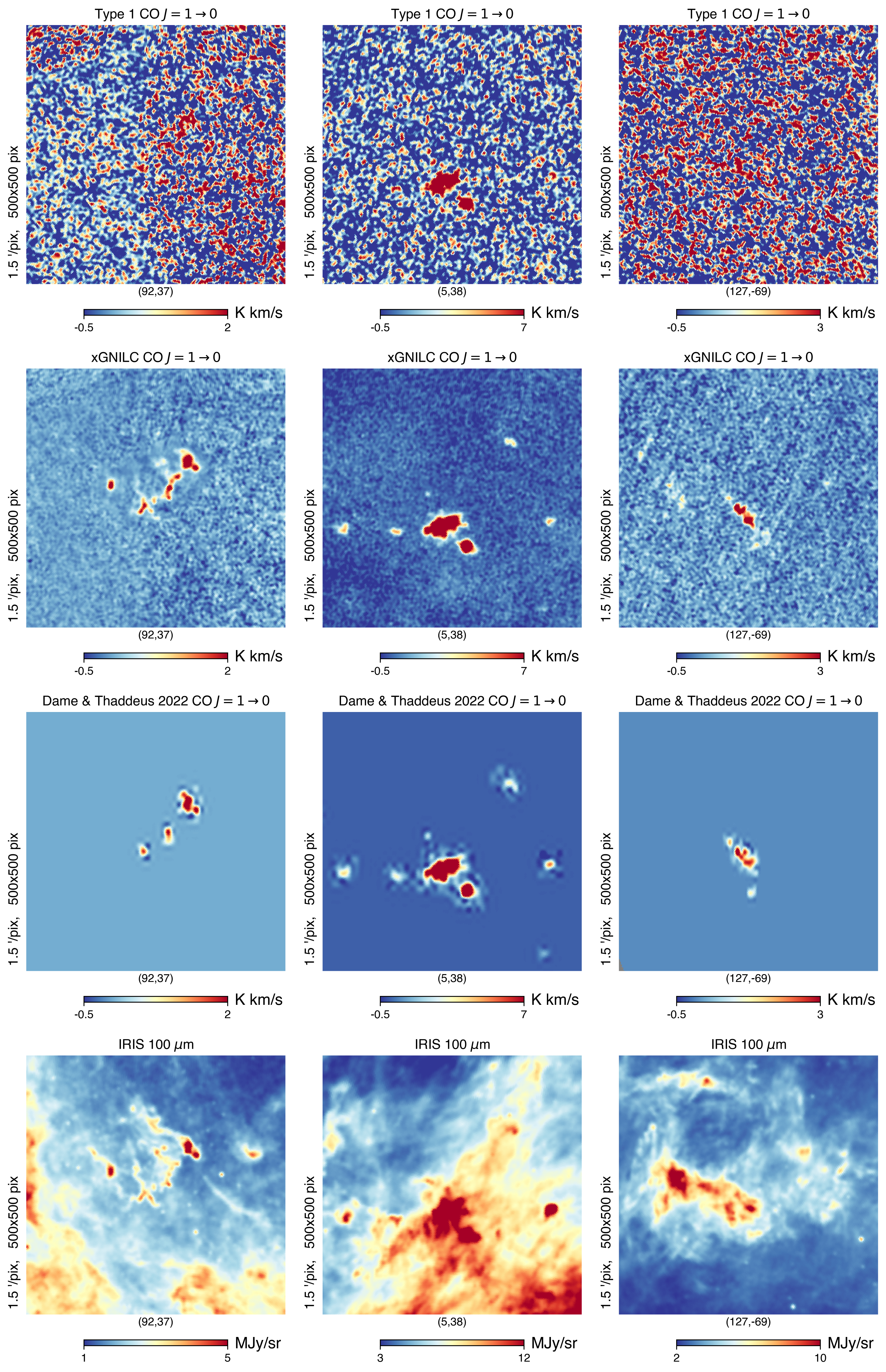}
    \caption{Images $(12.5^\circ \times 12.5^\circ)$ of few high latitude CO clouds and dust. The first row shows the Planck \typeone\ CO \coj\ map also smoothed with the same $10^\prime$ beam, second row shows our xGNILC CO \coj\ map, third row shows the \citet{2022ApJS..262....5D} map, and in the last row we show the IRIS 100 $\mu$m map smoothed to $10^\prime$ beam.}
    \label{fig:high-latitude_comparison}
\end{figure*}

We show the CO \cojjj\ map comparison in Fig. \ref{fig:co32_region_comparison} at $20^\prime$ resolution to reduce the noise in the Planck map. The signal-to-noise of the Planck map shown here is poor, particularly in the Polaris region. In the Taurus and Orion regions we find good agreement between the Planck \typeone\ and xGNILC. 
The difference maps for Taurus and Polaris are consistent with noise. The difference map in the Orion region shows small variations between the Planck \typeone\ and xGNILC, correlated with the signal, but this difference is small compared to the CO emission.

We also investigate high latitude Galactic CO emission seen in the xGNILC data products. The Planck \typeone\ CO maps are too noisy to contribute much of reconstructed CO signal at these high latitudes. Most of the reconstruction relies on the \typetwo\ maps and can have residual dust contamination. To trace the dust in regions of interest we use the IRAS 100 $\mu$m map \citep{IRIS:2005}. 

We focus on the north ecliptic sky for this study, to make use of the \citet{2022ApJS..262....5D} data. For this purpose we take the DT22+DHT masked datacube\footnote{\url{https://lweb.cfa.harvard.edu/rtdc/CO/NorthernSkySurvey/}} and integrate the cube over the velocity axis. The velocity integrated data has a beam of $\sim 8.4^\prime$ with a $0.25^\circ$ sampling. We interpolate the data to a $1^\prime$ grid and project to Healpix pixellization with $N_{\rm side}$ of 1024. Finally, we change the beam size from $8.4^\prime$ to $10^\prime$.

In Fig. \ref{fig:high-latitude_comparison} we show the comparison for three sample high latitude regions. We see good agreement between the xGNILC CO map and \citep{2022ApJS..262....5D}. All three regions have strong correlation with the dust but the xGNILC maps do not show any obvious excess dust contamination. In the Northern ecliptic sky where we have the new \citet{2022ApJS..262....5D} observations of faint CO clouds (excluding the Galactic region of the previous surveys), the xGNILC CO \coj\ map has a Pearson's correlation coefficient of $0.91$ for pixels brighter than 0.1 ${\rm K_{RJ}\, km\, s^{-1}}$. This indicates excellent agreement between the two observations, and gives confidence that the xGNILC CO maps do not suffer from very significant dust contamination. In Fig. \ref{fig:high-latitude_comparison} we also show the Planck \typeone\ CO \coj\ maps in the three regions. The noise levels of Planck \typeone\ maps make it difficult to see anything but the brightest CO clouds. Detecting the faint CO clouds at high Galactic latitude is significantly easier with the new xGNILC maps.

\begin{figure}
    \centering
    \includegraphics[width=0.37\textwidth]{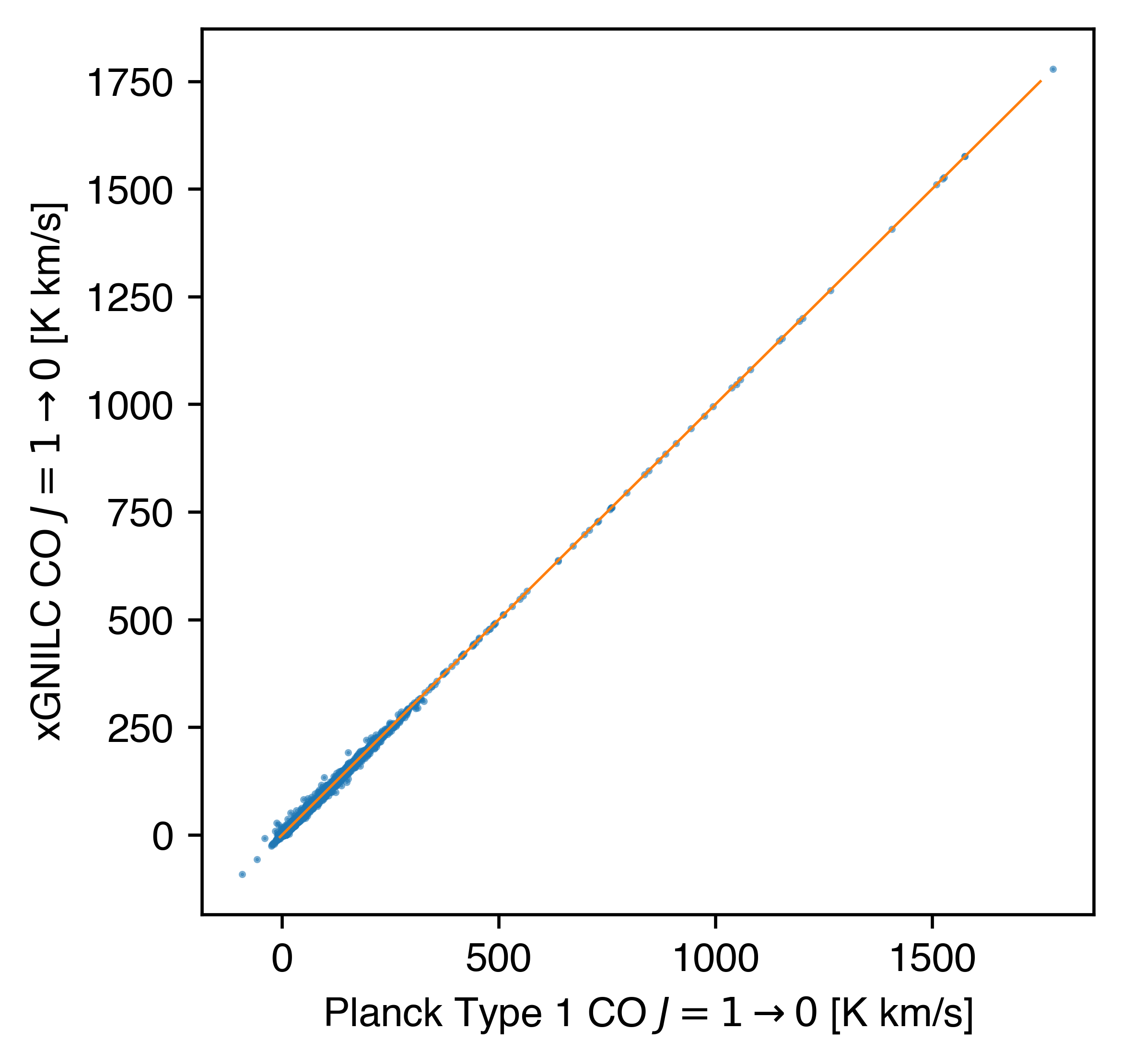}
    \includegraphics[width=0.37\textwidth]{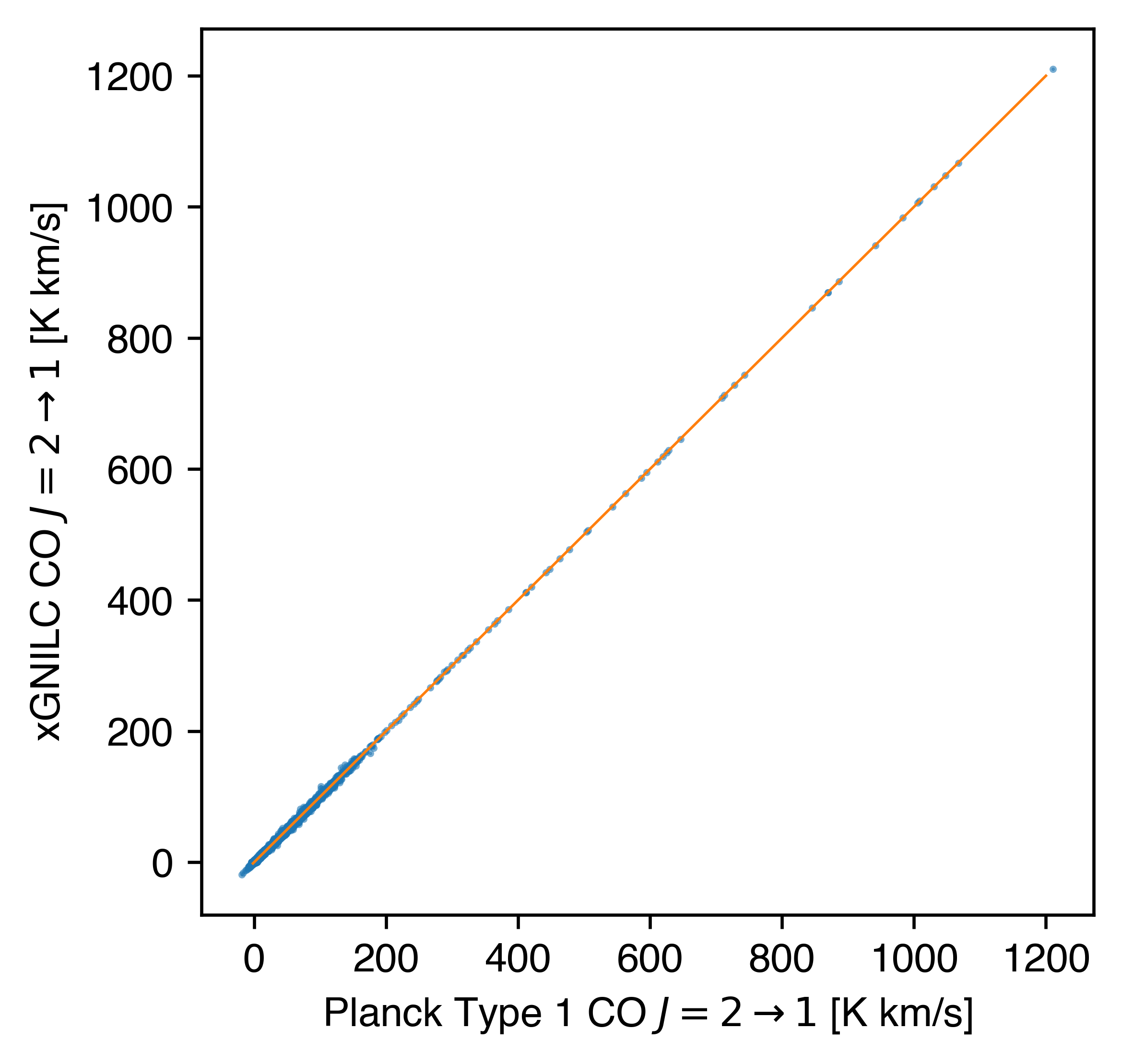}
    \includegraphics[width=0.37\textwidth]{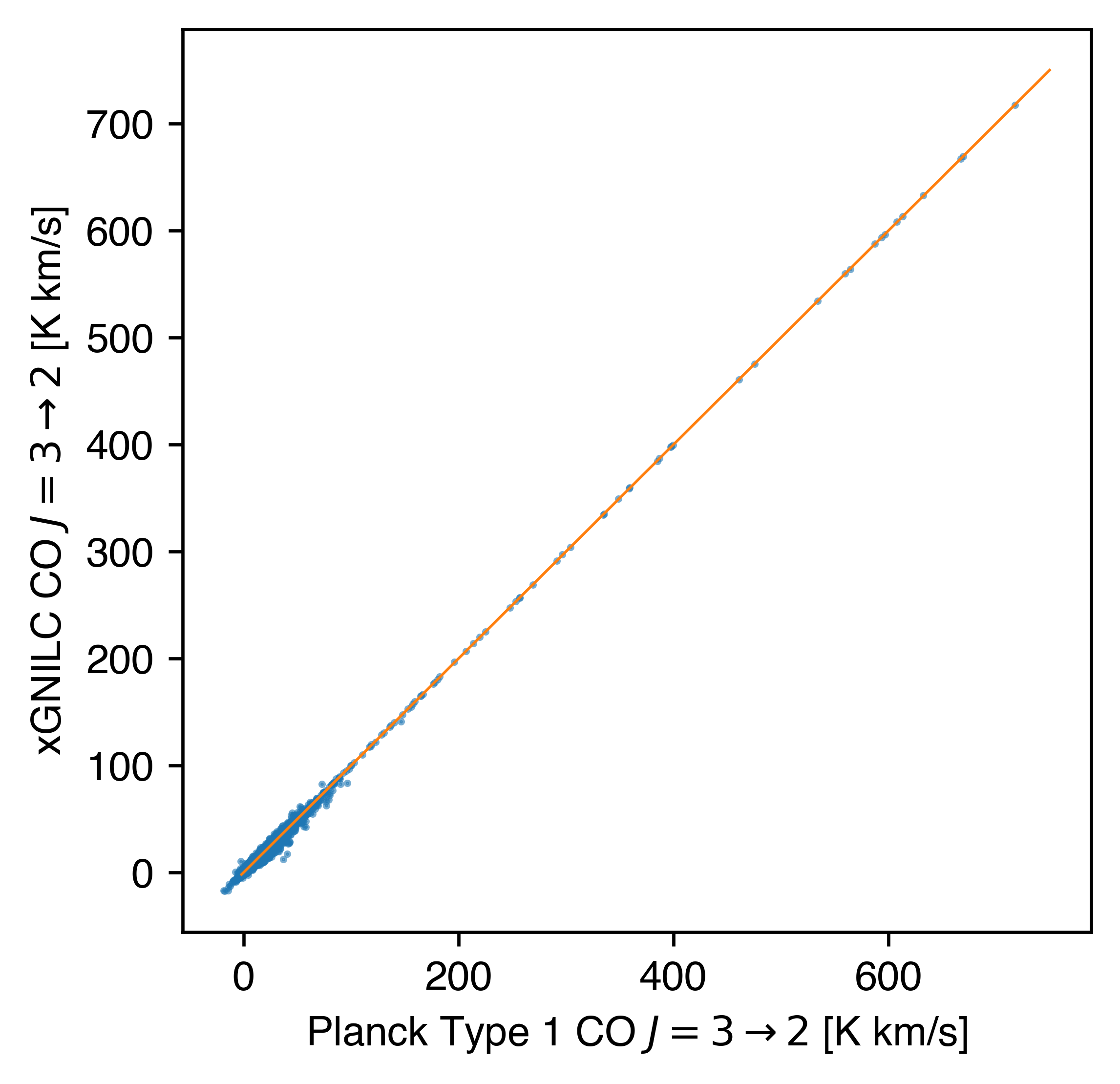}
    \caption{Scatter plot with pixel values of Planck \typeone\ map as abscissa and pixel values of xGNILC map as ordinate. Top: CO \coj; Middle: \cojj; Bottom: \cojjj.
    Planck \typeone\ maps are smoothed to $10^\prime$ resolution, and the HEALPix pixel size is increased to $N_{\rm side}=256$ for this figure. 
    The orange line indicates $y=x$ in all plots.}
    \label{fig:Ty1-xGNILC_scatter}
\end{figure}

In Fig. \ref{fig:Ty1-xGNILC_scatter} we plot a scatter of xGNILC map pixels with corresponding Planck \typeone\ map pixels at $10^\prime$ resolution, downgraded to HEALPix $N_{\rm side}=256$. Unlike similar plots for Planck \typetwo\ vs \typeone\ maps discussed in Sec. \ref{sec:limitations_of_existing_maps}, we see no disagreement between the xGNILC and Planck \typeone\ maps, and near-perfect correlation. At low pixel values we find slightly increased spread due to the noise in the Planck \typeone\ maps. 

\begin{figure}
    \centering
    \includegraphics[width=0.47\textwidth]{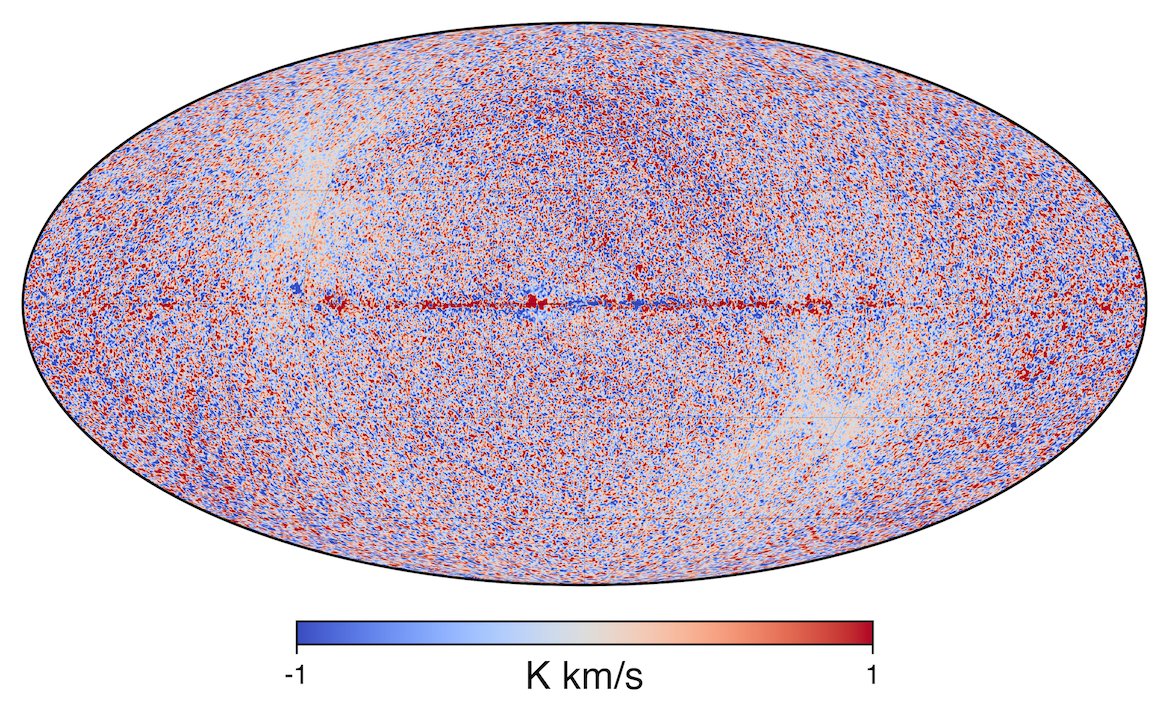}
    \includegraphics[width=0.47\textwidth]{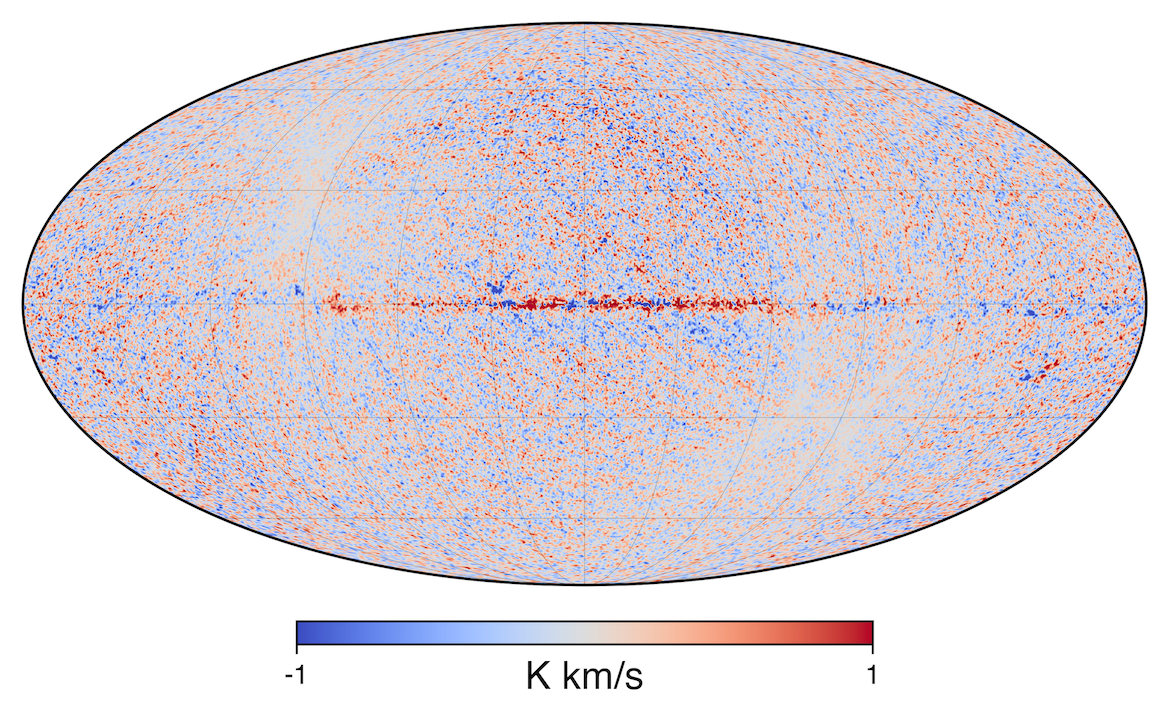}
    \includegraphics[width=0.47\textwidth]{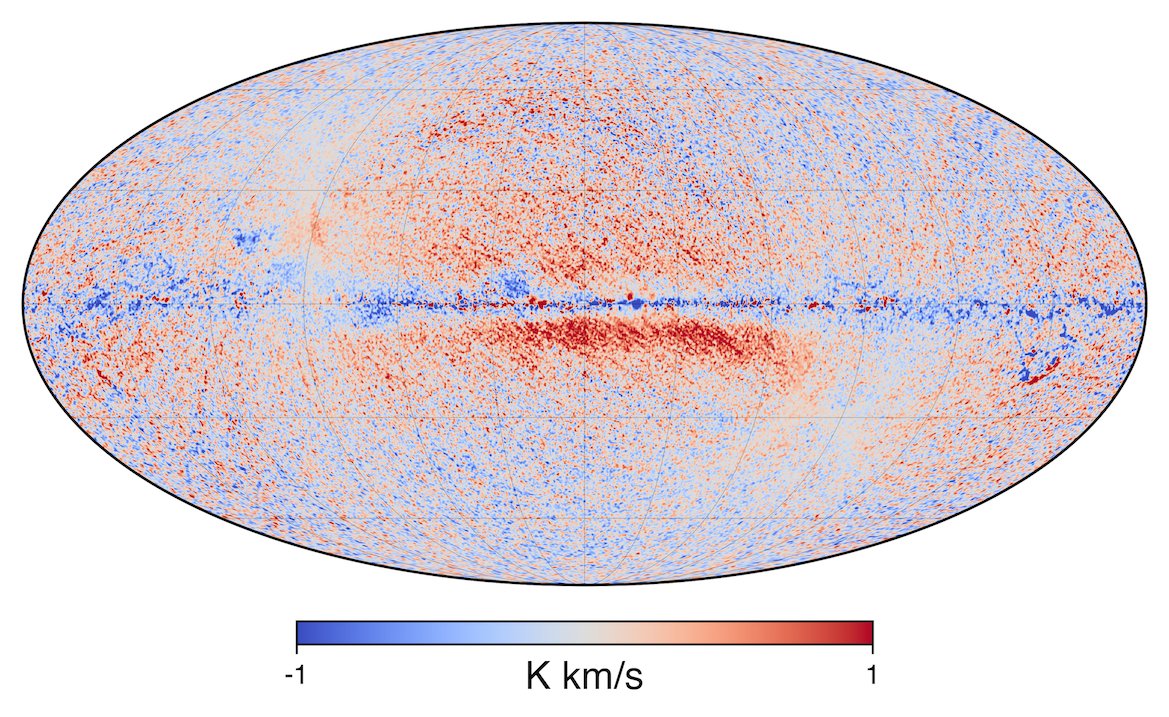}
    \caption{Difference maps (Planck \typeone\ $-$ xGNILC) for CO \coj\ (top), CO \cojj\ (middle), and CO \cojjj\ (bottom) are shown in this figure. Difference are visible in the Galactic plane, but are a small fraction of the total signal and lower than calibration uncertainties. For CO \cojjj, the effect of the postprocessing of residual systematics shows up as a negative trough close to the Galactic plane, and extended bright red region above and below the Galactic ridge.}
    \label{fig:Ty1-xGNILC_diff_maps}
\end{figure}

Next we take a look at the difference between the original Planck \typeone\ and the xGNILC CO maps. We smooth both maps to $30^\prime$ resolution, and then produce difference maps, shown in Fig. \ref{fig:Ty1-xGNILC_diff_maps}. For all three CO line emissions, we see small differences in the Galactic plane. In the case of CO \coj\, this difference is barely above the noise threshold, and for both CO \coj\ and \cojj\ emissions, the difference in the Galactic plane is small compared to the signal in the Galactic plane. In the case of CO \cojjj\ maps, the difference maps is largely a display of the systematic residuals that were visible in the Planck \typeone\ CO \cojjj\ map, and were subtracted by the postprocessing step described in Sec.~\ref{sec:postprocessing}. 


\subsection{Power spectrum validation}
\begin{figure}[t]
    \centering
    \includegraphics[width=0.45\textwidth]{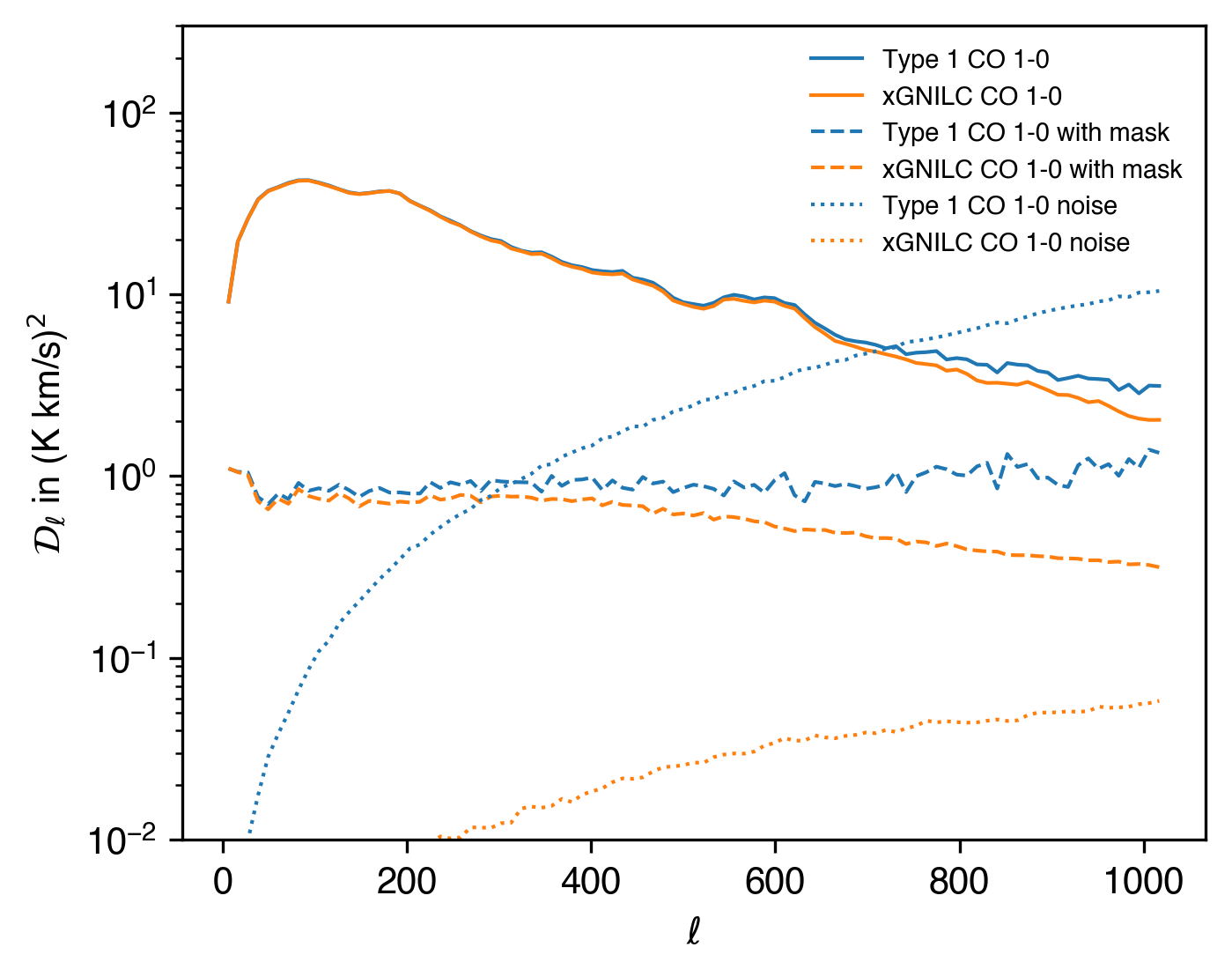}
    \includegraphics[width=0.45\textwidth]{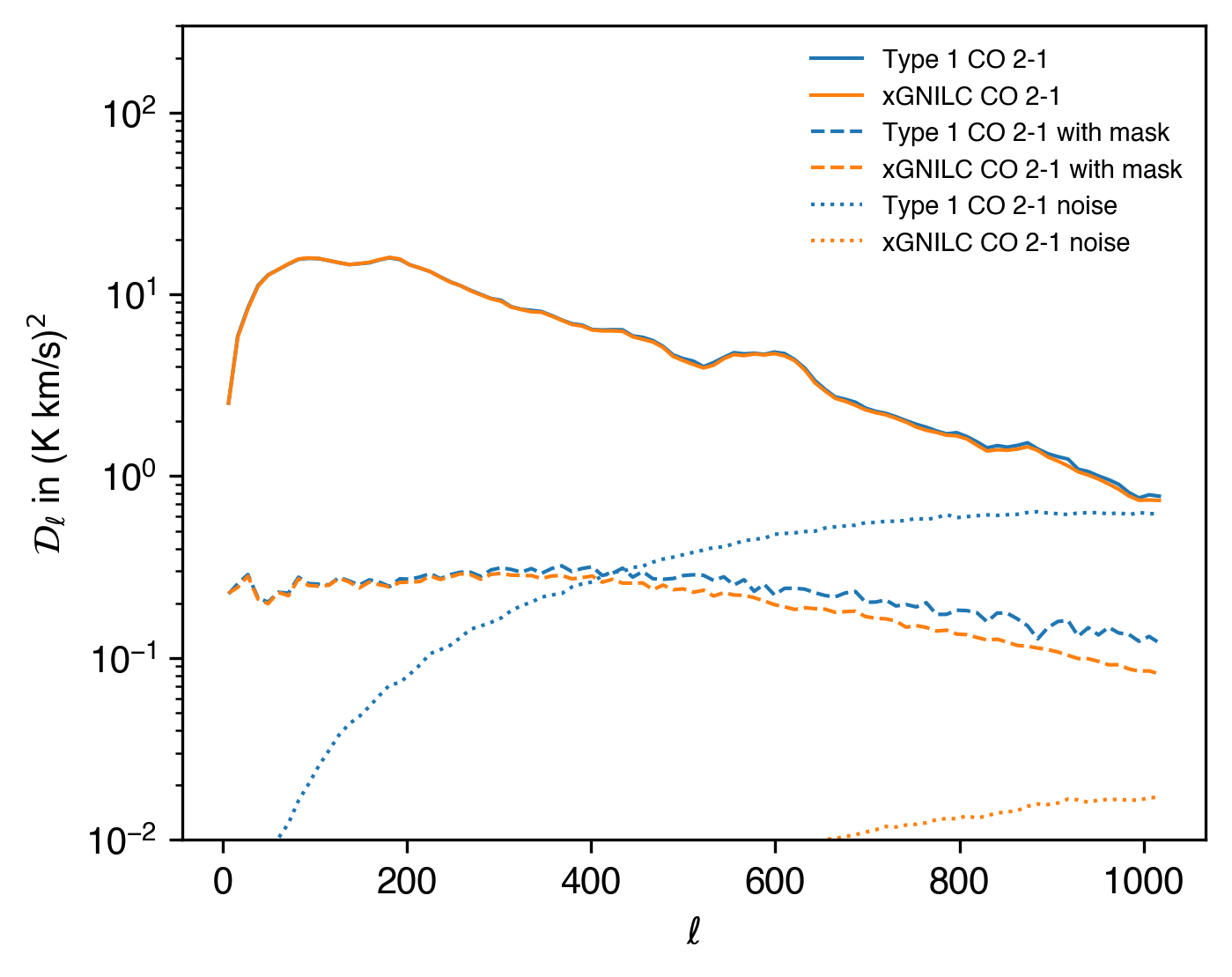}
    \includegraphics[width=0.45\textwidth]{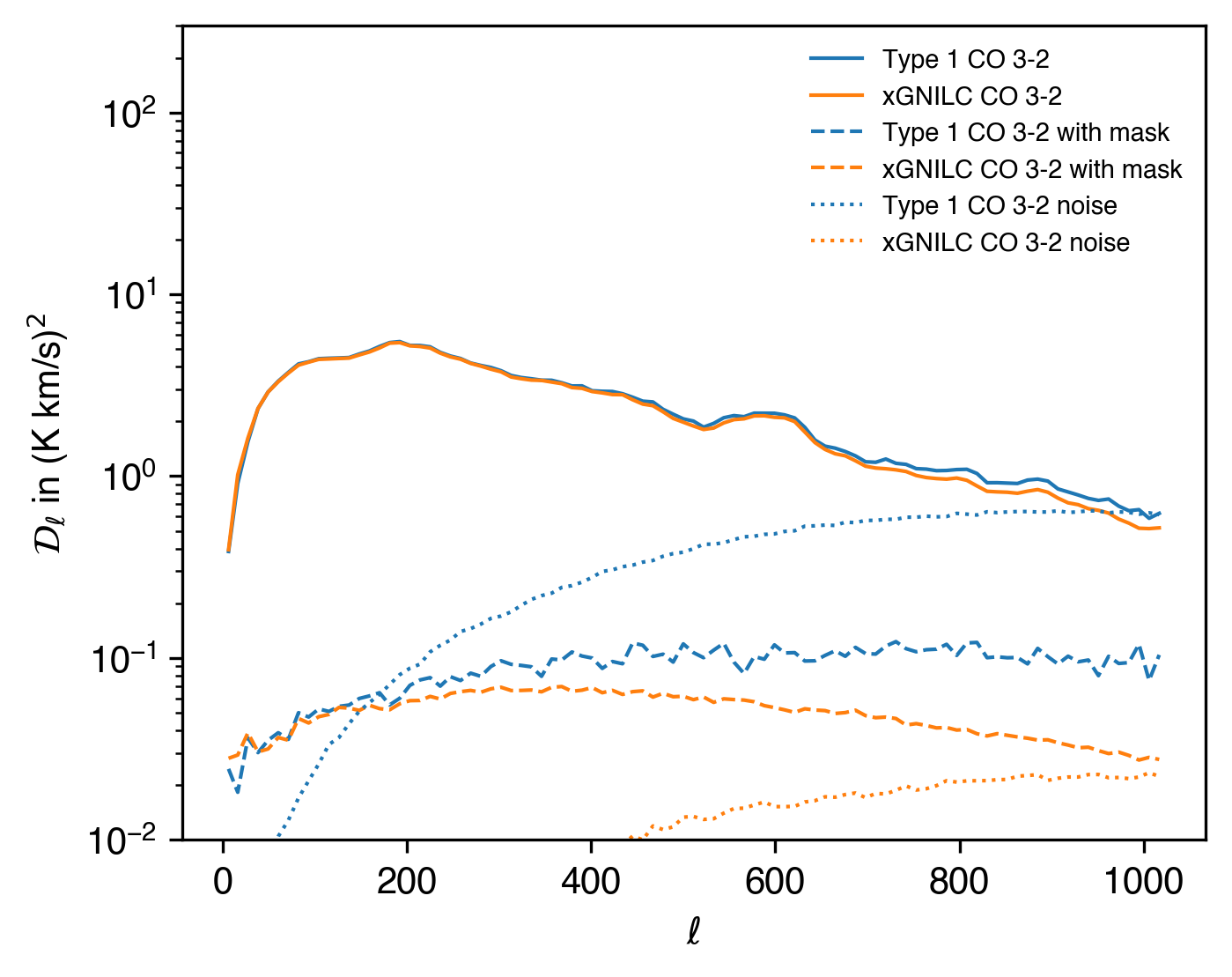}
    \caption{Power spectrum comparison for Planck \typeone\ CO maps (blue) and xGNILC CO maps (orange) at $10^\prime$ resolution. We noise debiased the \typeone\ map spectra shown in the figures. The top figure shows spectra comparison for the CO \coj\ maps, the middle figure for the CO \cojj\ maps and the bottom figure for the CO \cojjj\ maps.}
    \label{fig:power-spectra_validation_type1}
\end{figure}

\begin{figure}[t]
    \centering
    \includegraphics[width=0.45\textwidth]{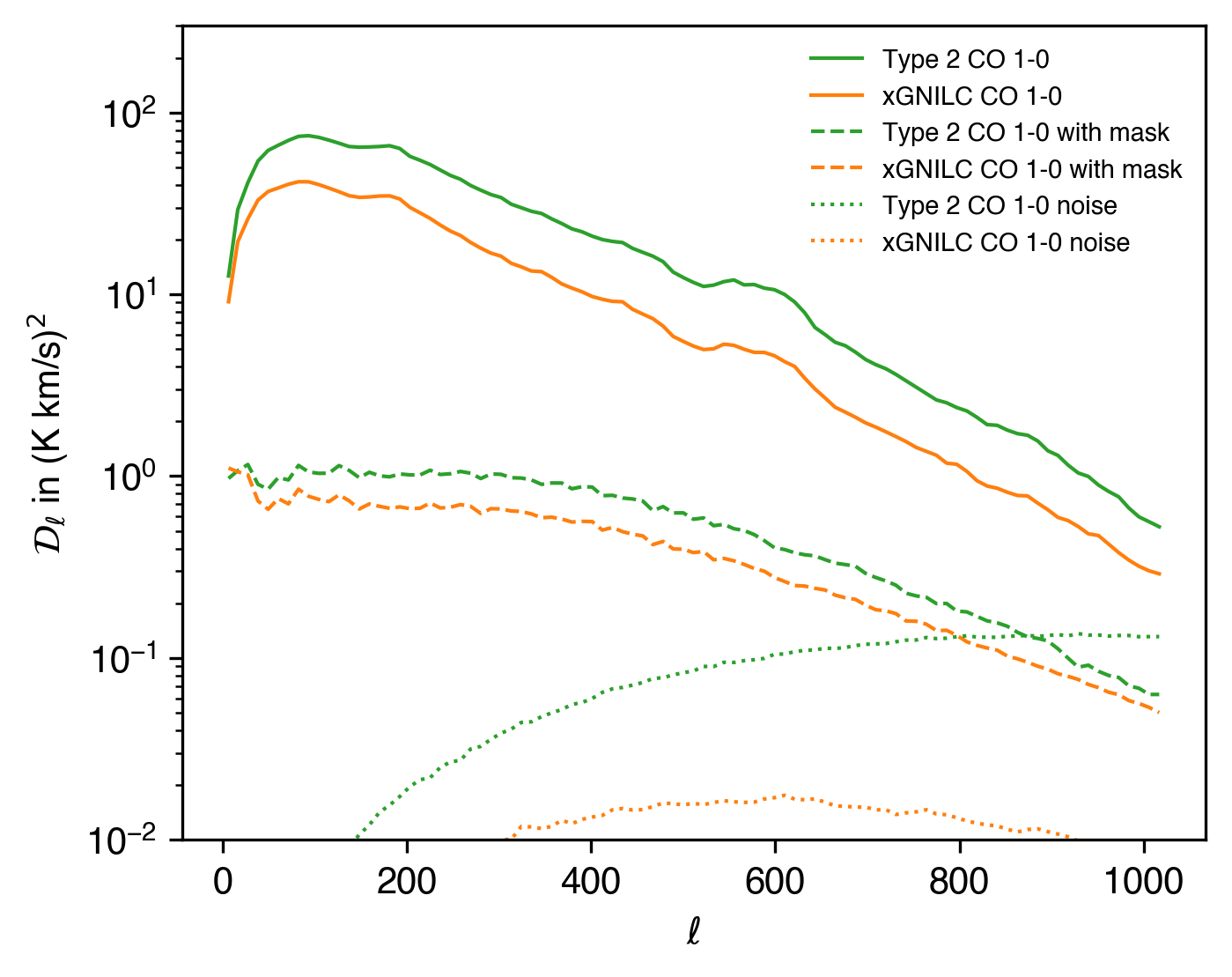}
    \includegraphics[width=0.45\textwidth]{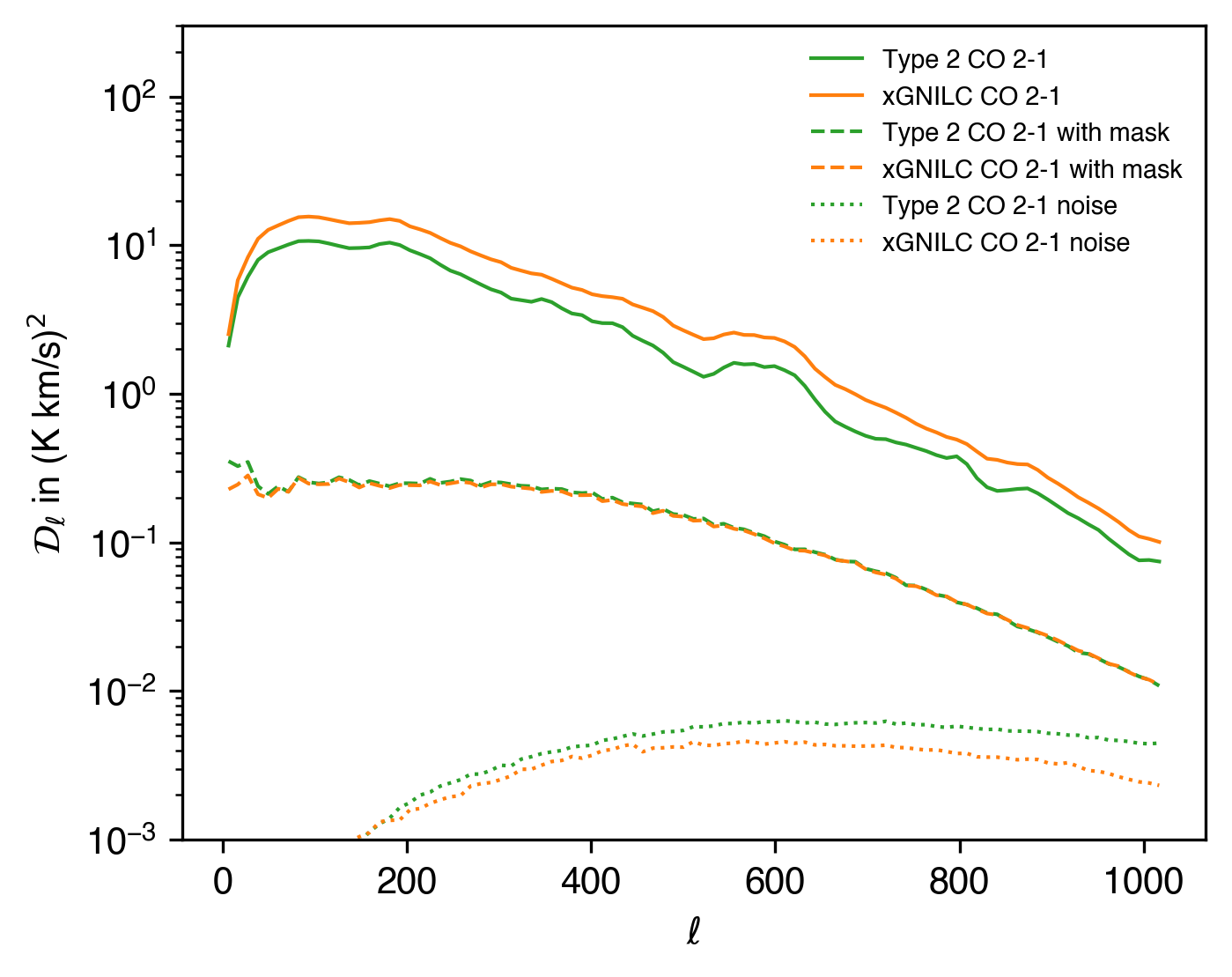}
    \caption{Power spectrum comparison for Planck \typetwo\ CO maps (green) and xGNILC CO maps (orange) at $15^\prime$ resolution. The top figure shows spectra comparison for the CO \coj\ maps and the bottom figure for the CO \cojj\ maps.}
    \label{fig:power-spectra_validation_type2}
\end{figure}

We now compare map power spectra (Fig. \ref{fig:power-spectra_validation_type1} and \ref{fig:power-spectra_validation_type2}).
For comparison of xGNILC with Planck \typeone, we smooth the \typeone\ maps to $10^\prime$, while  for comparison with the \typetwo\ maps we smooth the xGNILC maps to $15^\prime$. We use HEALPix \texttt{anafast} to compute the power spectra on the full sky, and also with the brightest regions masked in the Galactic ridge using the mask shown in Fig. \ref{fig:masks}. The spectrum is smoothed with a bin width of 7 multipoles to average out the fluctuations. The null maps are used to estimate the noise power spectra, and we show the \typeone\ and \typetwo\ power spectra after noise debiasing. There is no need to debias the power spectra for the xGNILC maps, as those are strongly signal-dominated over the full range of $\ell$ multipoles. 

Comparisons of xGNILC and Planck \typeone\ power spectra are shown in Fig. \ref{fig:power-spectra_validation_type1}. In the xGNILC maps, the signal dominates over the noise for the full $2 \le \ell \le 1000$ multipole range. An improvement by orders of magnitude of the noise level is achieved with the xGNILC postprocessing. 
Signal spectra for the xGNILC and Planck \typeone\ maps agree well where the \typeone\ spectra are above the noise, but deviate from one another at high $\ell$, where noise dominates. The power in xGNILC maps is lower than the power in the original maps after noise-debiasing. 
This is not unexpected, as even for the xGNILC algorithm (as compared to standard GNILC) some of the signal at small scales, where the input \typeone\ maps are low signal-to-noise, is filtered in the step of projection onto the signal subspace. It is plausible that in addition, null maps for the \typeone\ products do not capture all of the `noise' properties (and in particular contamination by point sources), and thus there is a small amount of correlated error in the half-ring maps, which contributes to the \typeone\ `signal' spectra but is at least partially filtered by xGNILC. For the CO \cojjj\ line emission specifically, systematic residuals present in the \typeone\ map have been removed in the xGNILC map, reducing also the total power on large scale.




We next compare the power spectra of the xGNILC maps with the Planck \typetwo\ maps (Fig. \ref{fig:power-spectra_validation_type2}). Differences seen in Fig. \ref{fig:ty1-ty2-spectra} between \typeone\ and \typetwo\ spectra are also present here, confirming that the signal in the xGNILC maps is strongly constrained to match the \typeone\ data products. The xGNILC CO \coj\ map has lower power than the corresponding  \typetwo\ map with and without masking, and the full sky power spectrum for CO \cojj\ xGNILC map is higher than the \typetwo\ map spectrum. However, the power spectra for the two \cojj\ maps with masking agree almost perfectly. This is expected because for the \cojj\ line calibration coefficients for the \typeone\ and \typetwo\ maps are almost the same, and the \typetwo\ CO \cojj\ map has the best signal-to-noise ratio. Hence, the GLS weights in the noisiest regions would basically prescribe dominant contribution from the \typetwo\ CO \cojj\ map.

In Fig.~\ref{fig:power-spectra_validation_type2}, we also note that the xGNILC CO \coj\ maps have lower noise than the corresponding \typetwo\ maps. This demonstrates that (as expected from the reprocessing pipeline by design) the noise level in the xGNILC maps is lower than in any of the original Planck CO maps, both \typeone\ and \typetwo.



\subsection{Limitations of the new maps}
In the previous two sections we validated our xGNILC data products by comparing them to the Planck CO data products. It is reasonable to argue that our maps are very high signal-to-noise CO maps that have most of the good features of the Planck \typeone\ CO data products. We reduce the contamination by noise, and improve the signal-to-noise on the small CO signal buried below the noise in the original \typeone\ maps. 
In addition, wherever there are discrepancies between \typeone\ and \typetwo\ maps in high signal-to-noise ratio regions, the xGNILC signal remains close to the original \typeone\ signal, and hence is presumably less contaminated by other Galactic foregrounds. xGNILC maps have also been cleaned from SZ clusters visible in \typetwo\ maps, as well as from residual systematic effects in the case of the \cojjj\ line.

In spite of these improvements, some limitations remain. Original calibration errors in the \typeone\ maps, as well as contributions from CO isotopologues, are basically unchanged. The \cojjj\ map, originally very noisy, is much cleaner after reprocessing, but this improvement is achieved in combination with maps of the other two lines, and some of the signal-correlated differences between the xGNILC and the \typeone\ maps, visible for instance in the Orion region, may be due to some level of line mixing. 

Our final data products combine \typeone\ and \typetwo\ maps.

One should note that strong correlation between the noise in the three line maps exist after the xGNILC reprocessing. 



\section{Conclusion}
\label{sec:conclusion}
In this work, existing Planck space mission maps of full sky rotational lines of Galactic CO emission have been revisited and post-processed to enhance signal-to-noise ratio and reduce  contamination by systematics and astrophysical confusion. Maps are produced at a common (`apodized') $10^\prime$ angular resolution. The data products include, in addition to the three CO line full-sky maps, the pipeline propagated null maps and maps of systematic uncertainty,
as well as processing and confidence masks. All data products are made available to the scientific community in a dedicated website.\footnote{Preliminary data release: \url{https://portal.nersc.gov/project/cmb/Planck_Revisited/co/}} 

\begin{acknowledgements}
The authors acknowledge useful discussions with Fran\c{c}ois Boulanger, Jean-Loup Puget, and Reijo Keskitalo. MR acknowledges support from the CSIC program \emph{Ayuda a la Incorporaci\'on de Cient\'ificos Titulares} provided under the project 202250I159. Some of the results in this paper have been derived using the HEALPix \citep{Gorski:2005} python package \citep{Zonca:2019}.
\end{acknowledgements}

\bibliographystyle{aa} 
\bibliography{references,Planck_bib} 


\end{document}